\documentclass[aps,prb,reprint,twocolumn,showpacs,superscriptaddress]{revtex4-2}

\usepackage{amsmath}
\usepackage{graphicx}
\usepackage{lmodern}
\usepackage{amsmath}
\usepackage{color}
\usepackage{amssymb}
\usepackage{bm}
\usepackage{braket}
\usepackage{mathtools}
\usepackage{afterpage}
\usepackage{soul}

\newcommand{\bk}{\bm{k}}
\newcommand{\bq}{\bm{q}}
\newcommand{\bp}{\bm{p}}
\newcommand{\bQ}{{\bm{Q}}}
\newcommand{\bP}{{\bm{P}}}

\renewcommand{\(}{\left(}
\renewcommand{\)}{\right)}
\renewcommand{\[}{\left[}
\renewcommand{\]}{\right]}

\newcommand{\be}{\begin{equation}}
\newcommand{\ee}{\end{equation}}

\usepackage[dvipsnames]{xcolor}
\usepackage[colorlinks=true]{hyperref}
\hypersetup{
    colorlinks=true,
    linkcolor=blue,
    citecolor=blue,
    urlcolor=blue
} 

\makeatletter
\def\maketitle{
\@author@finish
\title@column\titleblock@produce
\suppressfloats[t]}
\makeatother

\begin{document}
\title{$d$-wave Charge-$4e$ Superconductivity From Fluctuating  Pair Density Waves}
\author{Yi-Ming Wu}
\affiliation{Stanford Institute for Theoretical Physics, Stanford
  University, Stanford, California 94305, USA}
\author{Yuxuan Wang}
\email{yuxuan.wang@ufl.edu}
\affiliation{Department of Physics, University of Florida, Gainesville, Florida 32611, USA}

\date{\today}

\begin{abstract}
We present a theory for charge-$4e$ superconductivity as a leading low-temperature instability with a nontrivial $d$-wave symmetry.
We show that in several microscopic models for the pair-density-wave (PDW) state, when the PDW wave vectors connect special parts of the Fermi surface, the predominant interaction is in the bosonic pairing channel mediated by exchanging low-energy fermions. This bosonic pairing interaction is repulsive in the $s$-wave channel but attractive in the $d$-wave one, leading to a $d$-wave charge-$4e$ superconductor. By analyzing the Ginzburg-Landau free energy including {higher-order} fluctuation effects of PDW, we find that the charge-$4e$ superconductivity emerges as a vestigial order of PDW, and sets in via a first-order transition.
{Both the gap amplitude and the transition temperature decay monotonically with increasing superfluid stiffness of the PDW order.  
Our work provides a microscopic mechanism of higher-charge condensates with unconventional ordering symmetry in strongly-correlated materials.}
\end{abstract}
\maketitle

\noindent\textit{Introduction.---} 
\label{sec:introduction}
Charge-$4e$ superconductivity ($4e$-SC) is an exotic order in which four fermions are bound together and condense~\cite{PhysRevLett.80.3177,PhysRevLett.95.266404,PhysRevLett.94.247004,PhysRevB.82.134511,svistunov2015superfluid,li2022charge}, {with a nonvanishing order parameter $\sim\braket{\psi_i\psi_j\psi_k\psi_l}$ (where $\psi_j$ is some fermion field with a proper quantum number $j$)}. Compared with the more common charge-$2e$ superconductivity ($2e$-SC) {whose order parameter is $\sim\braket{\psi_i\psi_j}$}, it breaks the global $U(1)$ symmetry to a discrete $\mathbb{Z}_4$ symmetry, {in a sense that the discrete global gauge transformation $\psi_j\to\psi_j e^{in\pi/2}$ ($n=1,2,3,4$) leaves the order parameter invariant.}
{The fact that a $4e$-SC state is from binding four fermions can be reflected by the Little-Parks experiment, which was used as a direct proof of Cooper pairing phenomenon in $2e$-SC state. A recent such study was carried out on the Kagome metal CsV$_3$Sb$_5$~\cite{ge2022}, where a charge-$4e$ signal was observed in the normal state. However, whether one can observe the charge-$4e$ bounding in the superconducting state remains an active ongoing  research topic.}
{On the theory side, recent progress has been made in understanding the properties of a $4e$-SC~\cite{Gnezdilov,Jiang2017}, which is in general an nontrivial interacting system even at the mean-field level (unlike the Bogoliubov-de Gennes Hamiltonian of a $2e$-SC in which fermionic excitations are free).}
It was found~\cite{Gnezdilov} that $4e$-SC has a superfluid density perturbative in the order parameter, and is in general a gapless system.

However, the mechanism for $4e$-SC from interacting fermions remains a theoretical challenge. Unlike $2e$-SC which follows from an arbitrarily weak attractive interaction, there lacks a logarithmic divergence for the ``quadrupling" susceptibility for $4e$-SC.
{Some early speculations of inducing $4e$-SC by disorder~\cite{volovik1992exotic} or condensing skyrmions in a quadratic band touching system~\cite{PhysRevB.85.245123} exist. }
In recent years, a promising framework for understanding $4e$-SC has emerged from the perspective of intertwined orders~\cite{RevModPhys.87.457}, in which a plethora of orders breaking distinct symmetries can naturally emerge from the partial melting of certain primary electronic orders~\cite{Babaev2004,Agterberg2008,Berg2009,PhysRevLett.103.010404,Agterberg2011,Fernandes2021,Jian2021,liu2023charge}. 
The starting point of such analyses is typically a Ginzburg-Landau (GL) theory for the primary orders, and when the underlying fermionic models are taken into full account, the intrinsic leading instability toward $4e$-SC are in general subleading to those toward non-superconducting states, either nematic or ferromagnetic~\cite{Fernandes2021,https://doi.org/10.48550/arxiv.2301.01344,hecker2023cascade,hecker2023local}. It remains a major theoretical challenge to search for microscopic mechanisms for $4e$-SC as a \textit{leading} instability. 
Furthermore, in fermionic systems with repulsive interaction, $2e$-SC orders can emerge with unconventional pairing symmetries; while the pairing symmetry of $4e$-SC order has been largely unexplored~\cite{hecker2023cascade,liu2023charge} (see also Ref.~\cite{xiaoling2023} for an example in cold atom systems).

In this work,  we directly demonstrate that in a range of two-dimensional (2d) fermionic models with pair density wave (PDW) instabilities~\cite{Agterberg,FF,LO,cryst8070285,Matsuda2007,Gurevich,FeSC,doi:10.1126/science.abd4607}, a $4e$-SC order naturally emerges as the leading instability once fluctuation effects are taken into account. 
Extending the PDW flavor number to large-$M$ which justifies the mean field treatment of the $4e$-SC order, 
we establish the GL theory for the $4e$-SC state by incorporating higher order of interactions for the primary fluctuating PDW fields. 
(Of course, for  $M=1$, the phase transition in 2d is of Kosterlitz-Thouless nature, and the mean-field transition represents a crossover above the actual transition, but the transition is still driven by precisely the same attractive interaction we identify in this work.)
From this effective GL free energy we can determine the nature of $4e$-SC transition.  
We find the resulting  $4e$-SC order  has a $d$-wave symmetry, and the phase transition from high temperature normal state to $4e$-SC state is first-order. 
The transition occurs at $T_c>T_{\text{PDW}}$, where $T_{\text{PDW}}$ is the mean-field onset temperature of the primary PDW order. An important quantity that determines $T_c$ is the stiffness $\kappa$ of the PDW fluctuations, and 
we find that smaller stiffness  yields a higher $T_c$ as well as a larger gap amplitude $|\Delta_{4e}|$, as a larger $\kappa$ tends to suppress fluctuation effects and is hence detrimental for the development of composite orders. This may shed some new light on finding suitable platform materials for realizing the exotic $4e$-SC state. 

{\it Model and methods. $-$}
{The starting point of our theory, the PDW state, is by itself an exotic type of superconductivity with Cooper pairs carrying non-zero momenta.} {We emphasize that a microscopic mechanism for a PDW ground state in connection with realistic materials is not a settled issue. However, in recent years it} has been intensely studied and shown to emerge via various microscopic mechanisms~\cite{PhysRevLett.130.126001,PhysRevB.107.045122,PhysRevB.86.214514,PhysRevB.92.035153,https://doi.org/10.48550/arxiv.2209.14469,https://doi.org/10.48550/arxiv.2210.16324,PhysRevB.104.184501,Berg_2009,Zimmermann2011,PhysRevLett.114.197001,PhysRevLett.99.127003,PhysRevB.91.115103,PhysRevLett.114.197001,PALee2014,Nie2014,setty2021microscopic,setty2022exact,PhysRevLett.130.026001,Huang2022,Jiang2021,Li2023,Peng2021} to emerge. 
Regardless of the mechanism for PDW,
the instability toward $4e$-SC can be understood as the pairing instability due to interactions among PDW bosons. While in unconventional $2e$-SC the effective interaction among fermions stems from exchanging  bosons, here the effective interactions among bosons result from exchanging fermions, e.g., as illustrated in Fig.~\ref{fig:intro}(a).
{To make this effective interactions significant, the four fermions in Fig.\ref{fig:intro}(a) need to be as close to the Fermi surface as possible, and this can be realized by invoking a simple geometric relation in a $C_4$ symmetric system. } 

{In general, these PDW fluctuations can lead to different vestigial orders including CDW, nematic, and $4e$-SC that are of comparable strength (see, e.g., Refs~\cite{Fernandes2021,hecker2023cascade}).}
 However, we will show below that in a $C_4$ symmetric system, as long as the PDW momentum $\bQ$ satisfies 
\begin{align}
\frac{|\bm Q|}{\sqrt{2}} = \big|\bm k_F \!\!\!\!\mod (\pi,0)\big|,
\label{eq:0}
\end{align}
$4e$-SC order is naturally favored energetically over nematic and CDW orders. 
Here $\bm k_F$ {corresponds to one of the Fermi surface (FS) portions that are connected by the PDW order parameters, {i.e., the PDW hotspots}.}

{Notably, there are several microscopic theoretical models for PDW in which Eq.~\eqref{eq:0} is satisfied. For example, in Ref.~\cite{PhysRevLett.130.026001}, PDW order was shown to emerge from finite-range pair-hopping interactions, either in the continuum [illustrated in Fig.~\ref{fig:intro}(b)] or on a square lattice [Fig.~\ref{fig:intro}(c)]. In Refs.~\cite{PhysRevLett.114.197001,PhysRevB.91.115103}, PDW orders develop due to antiferromagnetic exchange interactions in the spin-fermion model proposed for the underdoped cuprates. Wile the applicability of these microscopic models to real materials remains an open issue, the ubiquity of the relation \eqref{eq:0} implies a generic formalism for $4e$-SC.}

Further, we stress  that no fine-tuning is needed{ --- we will see that even if Eq.~\eqref{eq:0} is approximately satisfied, our results will hold. In the continuum, all possible values of $\bm Q$ form a ring~\cite{PhysRevLett.130.026001}, and  the ratio $Q/k_F$ can be continuously varied. Moreover, we show in {the Supplementary Note I }~\cite{SM} that in several lattice models, this condition is \emph{automatically} satisfied for various values of $\bm k_F$ and $\bm Q$ that changes with doping.}
Below we show
that when Eq.\eqref{eq:0} is satisfied, the particular interaction in Fig.~\ref{fig:intro}(a) among {PDW bosons with momenta $\pm\bQ$ and $\pm \bP$ (wavevectors related by $\pi/2$ rotations) is parametrically enhanced.

The pairing interaction for PDW bosons are repulsive. We find that the repulsive interaction can lead to bosonic pairing ($4e$-SC) with uncoventional $d$-wave symmetry. We note that this is  reminiscent of the situation in fermionic pairing in unconventional superconductors. We develop a mean-field theory to capture the phase transition into the $4e$-SC phase. To that end, we introduced additional fields corresponding to bosonic bilinears, including the $4e$-SC order parameter, along with several Lagrange multiplier fields. After integrating out the auxilary fields, we obtain a free energy for $4e$-SC, and find that the phase transition is of first order.

\begin{figure}
  \includegraphics[width=8cm]{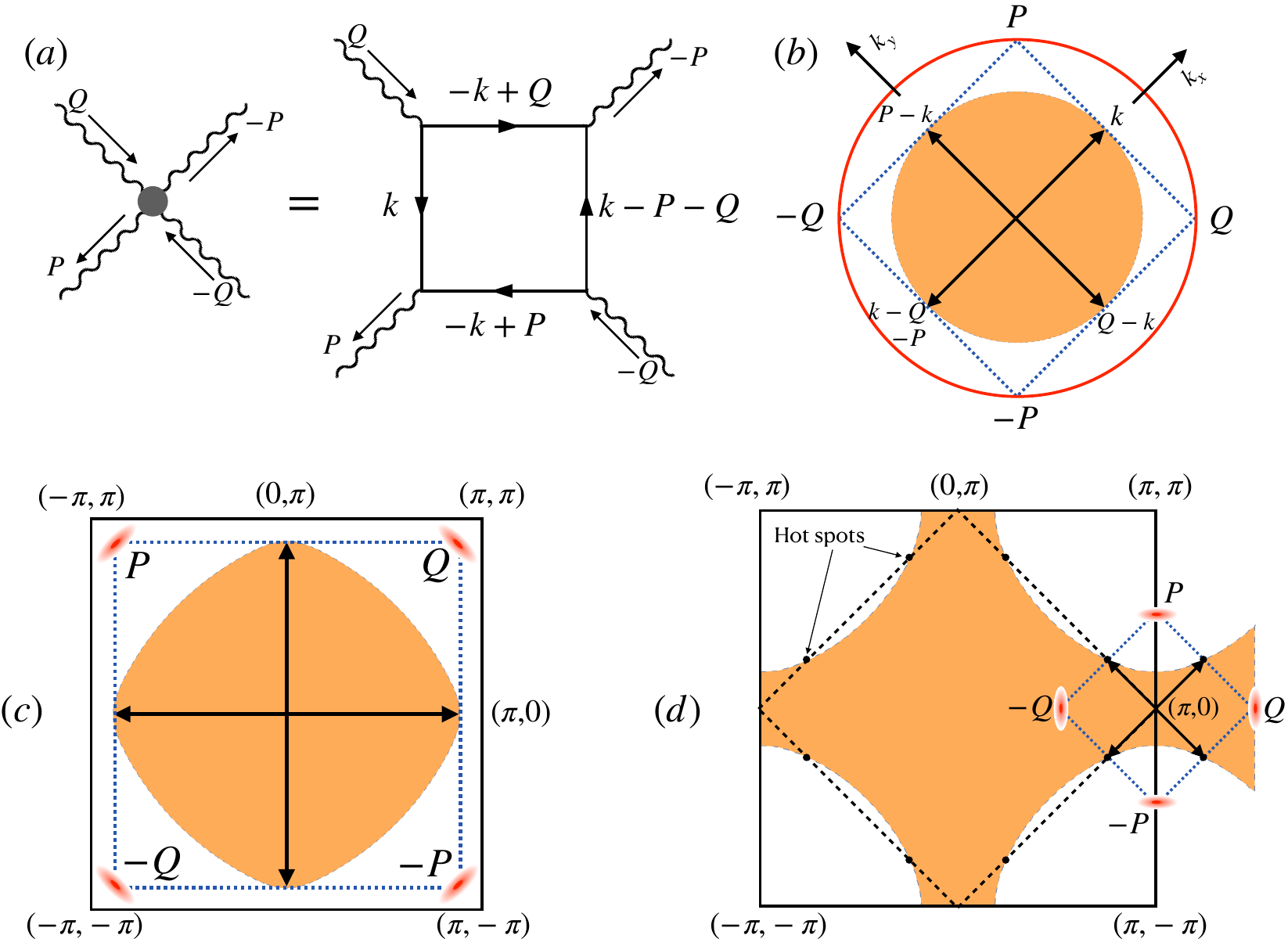}
  \caption{{$4e$-SC from fluctuating PDW.} (a) The effective pairing interaction for PDW bosons via exchanging fermions. Here $\pm \bQ$ and $\pm\bP$ are related by consecutive $\pi/2$ rotations. When Eq.~\eqref{eq:0} is satisfied, the internal fermions are on the Fermi surface. (b) Continuum model for PDW with $O(2)$ symmetry from Ref.~\cite{PhysRevLett.130.026001}. The orange area is the filled Fermi sea, and the PDW momentum $\bQ$ takes values on the Bose surface represented by the red circle. The model can be continuously tuned such that $|\bQ|=\sqrt{2}k_F$. (c) The PDW model realized on a square lattice when the fermions relevant for pairing are close to the Van Hove points $(0,\pi)$ and $(\pi,0)$ and the PDW momentum is close to $(\pi,\pi)$. (d) PDW flucutations in the spin-fermion model~\cite{PhysRevLett.114.197001,PhysRevB.91.115103}. The four hot spots (black dots) near $(\pi,0)$ form a square, and the momenta for the potential PDW orders also form a square. The models in (c) and (d) satisfy the condition in Eq.~\eqref{eq:0}.
  }\label{fig:intro}
\end{figure}

\noindent\textit{Attractive interaction from fluctuating PDW.---} As a starting point, we consider a model for fluctuating PDW order with a $C_4$ symmetry. The analysis for the case with continuous rotation symmetry will be qualitatively similar.
Denoting the bosonic PDW field by $\Psi_i(\bq)$, where $\bq$ is measured from the ordering momentum $\bQ_i\in\{\pm\bP,\pm\bQ\}$, 
the generic free energy {up to quadratic order} for PDW fluctuations of those examples depicted in Fig.~\ref{fig:intro} can be written as
\be
\label{eq:3}
\!\!\!\!\!F= \sum_{i=1}^4 \int\frac{d^2\bm q}{4\pi^2} \[\alpha(T) + \kappa\bm q^2\] \left | \Psi_i(\bm q)\right |^2 + \mathcal{O}(\Psi^4)
\ee
where $\kappa$ is the {amplitude} stiffness for the PDW orders {obtained by expanding the particle-particle susceptibility to $\mathcal{O}(\bq^2)$}, and below the mean-field ~\eqref{eq:6} below, this hierarchy allows the $4e$ interaction to be dominant over other channels.}
In general, the dispersion of the PDW bosons is anisotropic in $\bq$, but {as can be directly checked, for $4e$-SC instabilities the anisotropy factor can be absorbed into a redefinition of $q_x$ and $q_y$} around each PDW ordering momenta {(see the Supplementary Note III~\cite{SM})}. For PDW fluctuations intrinsically driven by finite-range electronic interactions, $\kappa$ comes from both the particle-particle bubble and from the $\bm q$ dependence of the interaction, and for weak interactions we assume that $1/\kappa \ll {E_F}$. {Further, as will be explained later, for analytical control of the theory, we are interested in the regime $1/\kappa \ll T_{\rm PDW}\ll {E_F}$. This corresponds to a regime in which PDW and $4e$-SC develop at low energies, and the phase fluctuations of PDW are not ``too strong"; see discussions around Eq.~\eqref{eq:6} and Eq.~\eqref{eq:112} below.} Indeed, in the microscopic theory of Ref.~\cite{PhysRevLett.130.026001}, $\kappa T_{\rm PDW}$ can be freely tuned by the interaction strength. Hereafter we measure all lengths against $n^{-1/2}$, where $n$ is electron density, and all energies against the Fermi energy, which, is defined as $E_F\equiv v_F\sqrt{n}$. 
 
The effective interaction between the PDW bosons {starts at the order of $|\Psi|^4$. It can be viewed as arising from exchanging fermionic degrees of freedom [see Fig.\ref{fig:intro}(a)]}. As is usually done for itinerant fermions, the processes involved are described by square diagrams. The key insight here is that, for dominant four-boson interactions, the internal fermions need to come from the vicinity of the FSs. {For systems satisfying Eq.\eqref{eq:0}}, this consideration alone singles out three types of interactions
\be
\label{eq:4}
\beta_1|\Psi_0|^4 + \beta_2 |\Psi_{\pm \bP}|^2|\Psi_{\pm\bQ}|^2 + \beta \(\Psi_\bP \Psi_{-\bP} \Psi^*_\bQ \Psi^*_{-\bQ}+h.c.\),
\ee
where we have used the shorthands, e.g., $|\Psi_0|^4 \equiv \sum_{i}\int_{\bq_1,\bq_2,\bq_3}\Psi_i(\bq_1) \Psi_i^*(\bq_2) \Psi_i(\bq_3) \Psi_i^*(\bq_1-\bq_2+\bq_3)$, and $\int_{\bq}\equiv \int\frac{d^2\bq}{4\pi^2}$. Importantly, the last interaction with coefficient $\beta$, depicted diagrammatically in Fig.~\ref{fig:intro}(a), is of appreciable strength only when the condition Eq.~\eqref{eq:0} is satisfied; otherwise at least two fermion propagators would come from regions far away from the FS. 
The coefficients $\beta_{1,2}$ and $\beta$ can be readily obtained by integrating out low-energy fermions, and by linearizing the fermionic dispersion we obtain {(see Ref.~\cite{PhysRevB.90.035149} and also the Supplementary Note II \cite{SM} for details)} 
\be
\label{eq:6}
\beta_1 \sim 1\textrm{, }\beta_2 \sim  \ln\frac{1}{T}\textrm{, and }\beta = \frac{1}{16 T}.
\ee 
 Observe that  parametrically, $\beta \gg \beta_{1,2}$ as long as $T\ll 1$ (in original units, $T\ll E_F$). {Due to the large separation between $\beta$ and $\beta_{1,2}$, we expect that even if Eq.~\eqref{eq:0} hold approximately, one can safely neglect $\beta_{1,2}$.} 
 
 The last term in Eq.~\eqref{eq:4} represents a repulsion  for the bosons corresponding to PDW fluctuations. As is familiar from $2e$-SC, repulsive interactions may have attractive components in pairing channels with higher-angular momenta. To this end,  we introduce bilinear operators $\Phi_s$ and $\Phi_d$ for $4e$-SC with $s$-wave and $d$-wave components
  \begin{equation}
  	\Phi_{s/d} (\bq)\equiv \int_{\bm p} \Psi_{\bQ}(\bm p+\bq)\Psi_{-\bQ}(-\bm p)\pm\Psi_{\bP}(\bm p+\bq)\Psi_{-\bP}(-\bm p)
  \end{equation}
 such that the $\beta$ interaction can be rewritten as
\begin{equation}
\frac{\beta}{2} \int  \frac{d^2 \bq }{4\pi^2}|\Phi_s(\bq)|^2-\frac{\beta}{2}\int  \frac{d^2 \bq }{4\pi^2}|\Phi_d(\bq)|^2.\label{eq:beta_dwave}
\end{equation}
Just like their $2e$-SC counterparts, $\Phi_{s,d}$ are distinguished by their transformation properties under the $C_4$ rotation: $\Phi_s$ is even and $\Phi_d$ is odd. 
From this decomposition  it is evident that in $d$-wave channel the charge-$4e$ pairing interaction is attractive. 

Let us comment on other couplings and fluctuations in other channels. The $\beta_1$ term corresponds to a local repulsive interaction between the PDW bosons. The $\beta_2$ term is repulsive, but it can be written as ${\beta_2}\(|\Psi_{\bP}|^2+|\Psi_{\bQ}|^2\)^2/4 - {\beta_2}\(|\Psi_{\bP}|^2-|\Psi_{\bQ}|^2\)^2/4$,
revealing its tendency toward a nematic instability with the order parameter 
$\mathcal{N} \sim |\Psi_{\bP}|^2-|\Psi_{\bQ}|^2$.
{In several recent works~\cite{Fernandes2021,hecker2023cascade} based on two uniform superconducting orders, it was generally found that the vestigial nematic order is more favorable. However, our microscopic calculation has shown that in our PDW-based model, thanks to the $C_4$ symmetry and the condition in Eq.\eqref{eq:0}, $4e$-SC is favored energetically, as $\beta\gg \beta_2$.}

We also note that the decomposition of the $\beta$-term interaction is not unique. Notably, it can also be decomposed such that there is an equally attractive interaction for a charge-density-wave (CDW) composite order, with e.g., $\rho_{\bP-\bQ} = \Psi_\bP \Psi^*_\bQ -  \Psi_{-\bQ} \Psi^*_{-\bP}$. The interplay between CDW and $4e$-SC has been systematically studied in a phenomenological model~\cite{Berg2009}. However, in our microscopic model the CDW instabilities are secondary to that of $4e$-SC. Qualitatively, the reason is similar to the situation in a fermionic theory --- in 2d and higher dimensions with weakly-coupled fermions, the CDW instability requires nesting of the FS, that is, fermionic dispersions at different momenta with a fixed  difference need to be the same, while the $2e$-SC instability 
{does not}.~\cite{PhysRevB.90.035149}
 Here the same reasoning holds for interacting bosons, where the  bosonic dispersion for $\Psi_{\bP}$ and $\Psi_\bQ$ is clearly not nested, which can be directly seen from Fig.~\ref{fig:intro}.
Therefore, the CDW is suppressed compared with the $d$-wave $4e$-SC, even if  attractive interactions for these channels are equal {(see {the Supplementary Note V}~\cite{SM} for a quantitative analysis)}. 

Just like the quartic interactions, higher-order interactions come microscopically from higher-order processes of exchanging low-energy fermions. Making use of the condition in Eq.~\eqref{eq:0}, again only a subset of diagrams need to be included. We present all the dominant interactions at sixth and eighth order in {the Supplementary Note II} ~\cite{SM}, and find that the sixth-order coupling constant $\gamma$ and eighth-order coupling constant  are given by $\gamma={1}/({768T^3)}$, and $\zeta={1}/({7680 T^5})$.

\noindent\textit{Formation of $d$-wave $4e$-SC.---}
We obtain a phase transition into a $d$-wave $4e$-SC, conspired by  fluctuations of the PDW bosons described by Eq.~\eqref{eq:3}, and the attractive interaction in Eq.~\eqref{eq:beta_dwave}. Formally, the mean-field theory can be justified by extending  $\Phi_i(\bq)$ to an $M$-component field $\Phi_i^J(\bq)$ where $J\in[1,M]$~\cite{SM,PhysRevB.85.024534,PhysRevB.90.035149}.

We first take the mean-field ansatze $\mathcal{N}=\Phi_s=0$. We then have up to eighth order in $\Psi$ {(see details in the Supplementary Note III and IV \cite{SM})}
 \begin{align}
    {F}=& \sum_{i} \int_{\bq} \[\alpha(T) + \kappa\bm q^2\] \left | \Psi_i(\bm q)\right |^2 -\frac{\beta}{2}\int_{\bq}|\Phi_d(\bq)|^2 \nonumber\\
  &  -\frac{\gamma}{4}\int_{\bq,\bp}\Phi_d(\bq)\Phi_d^*(\bq+\bp)\phi(\bp)    \label{eq:15}\\
  &+\frac{\zeta}{16}\int_{\bq,\bk, \bp}\Phi_d(\bq)\Phi_d^*(\bq+\bp)\Phi_d(\bk)\Phi_d^*(\bk-\bp)
  \nonumber\\
    &-\frac{\zeta}{{16}}\int_{\bq,\bk, \bp}\Phi_d(\bq)\Phi_d^*(\bq+\bp)\phi(\bk)\phi^*(\bk-\bp)+ \mathcal{O}(\Psi^{10}),\nonumber
  \end{align}
  where $\phi\equiv |\Psi_{\bP}|^2+|\Psi_{-\bP}|^2+|\Psi_{\bQ}|^2+|\Psi_{-\bQ}|^2$ is the Gaussian fluctuation.

The decoupling of the interaction terms  of $\Psi$ at any order can be achieved by introducing \textit{two} sets of ancillary fields $\lambda_d,\mu$ and $\Delta_d,\varphi$, where $\lambda_d$ and $\mu$ are Lagrange multiplier fields ensuring $\Delta_d=\Phi_d$ and $\varphi=\phi$.~\cite{SM} We can then replace the bilinear operators $\phi$ and $\Phi_d$ with local fields. For quartic interactions we show in {the Supplementary Note VI }~\cite{SM} that the result is the same as that from the HS transformation.
 As a mean-field ansatz, we consider spatially uniform saddle-point solutions, with $\Delta_d(\bm q) = \Delta_{d}\delta (\bm q)$, and $\varphi(\bq) = \varphi \delta(\bq)$. 
Using the regularization $\delta(\bq=0)=V$ where $V$ is the volume (area) of the system and defining the free energy density $\mathcal{F}\equiv F/V$, we have
\begin{align}
    &\mathcal{F}[\Psi, \Delta_d, \varphi, \lambda_d, \mu ]= \sum_{i,\bq} \[r+\kappa\bm q^2\] \left | \Psi_i(\bm q)\right |^2 \nonumber\\
  &-\frac{\beta}{2}|\Delta_d|^2 +\frac{\zeta}{16}|\Delta_d|^4
-\frac{\zeta}{{16}}|\Delta_d|^2\varphi^2\nonumber\\
&+\lambda_d\({\Delta}_d^*-\Phi^*_d\)+\lambda_d^*\({\Delta}_d-\Phi_d\)+\mu\varphi+ \mathcal{O}(\Psi^{10}),
    \label{eq:16}
  \end{align}
  where $r=\alpha(T) -\frac{\gamma}{4}|\Delta_d|^2-\mu$, $\lambda_d\equiv \lambda_d(\bq =0)/V$, $\mu \equiv \mu(\bq =0)/V$, and we have used the fact that ${V}\int_{\bq}\equiv \sum_{\bq}$ in the continuum limit.  We note that $\mathcal{F}[\Psi, \Delta_d, \varphi, \lambda_d, \mu )$ is at most quadratic in the field $\Psi(\bq)$.
  
Assuming $r>0$, we can  integrate out the PDW fields and get a free energy density $\mathcal{F}( \Delta_d, \varphi, \lambda_d, \mu )$. Generalizing to large-$M$ and taking the saddle points for $\varphi, \lambda_d, \mu$, we get up to constant terms
\begin{equation}
	{\mathcal{F}( \Delta_d)}/{M} = A(T) |\Delta_d|^2+B(T) |\Delta_d|^4  + C(T)|\Delta_d|^6 + \cdots,
	\label{eq:111}
\end{equation}
{where the GL coefficients $A(T), B(T)$ and $C(T)$ are determined by the microscopic parameters $\kappa$, $\alpha, \beta, \gamma$ and $\zeta$ and their full expressions are listed in { the Supplementary Note IV} ~\cite{SM}. One important feature of these coefficients is each of them contains terms that are organized in increasing powers of $1/\kappa T\ll 1$ or $1/\kappa\alpha$. Assuming $\kappa\alpha\gtrsim 1$ as we will justify below, the coefficients can be greatly simplified by keeping the leading contributions,
  \begin{align}
    A(T)&\approx \frac{2 \pi  \alpha  \kappa }{T}-\frac{\beta }{2},~B(T)\approx-\frac{4 \pi ^3 \alpha  \kappa ^3}{3 T^3}, \nonumber \\
C(T)&\approx \frac{64 \pi ^5 \alpha  \kappa ^5}{45 T^5}
\label{eq:112}
  \end{align}}

Driven by the temperature dependence of $\alpha(T)$, the quadratic coefficient $A(T)$  becomes negative upon lowering temperature when $\alpha \kappa = \beta T/4\pi\approx0.005$. At this point, we see that $B(T)<0$, i.e., $\Delta_d=0$ is no longer a local minimum of the free energy. Instead, the global minimum of free energy is given by $\Delta_d\neq 0$. Therefore, a first-order phase transition must have already occurred at a higher temperature. {The transition temperature is thus set by 
\be
\label{eq:tc}
\alpha_c\sim 1/\kappa,~~~T_{c} >T_{\rm PDW},\
\ee
which can be directly confirmed by minimizing \eqref{eq:111} up to sixth order.
We see that a smaller stiffness term $\kappa$ leads to a higher $T_c$.}

\begin{figure}
  \includegraphics[width=8cm]{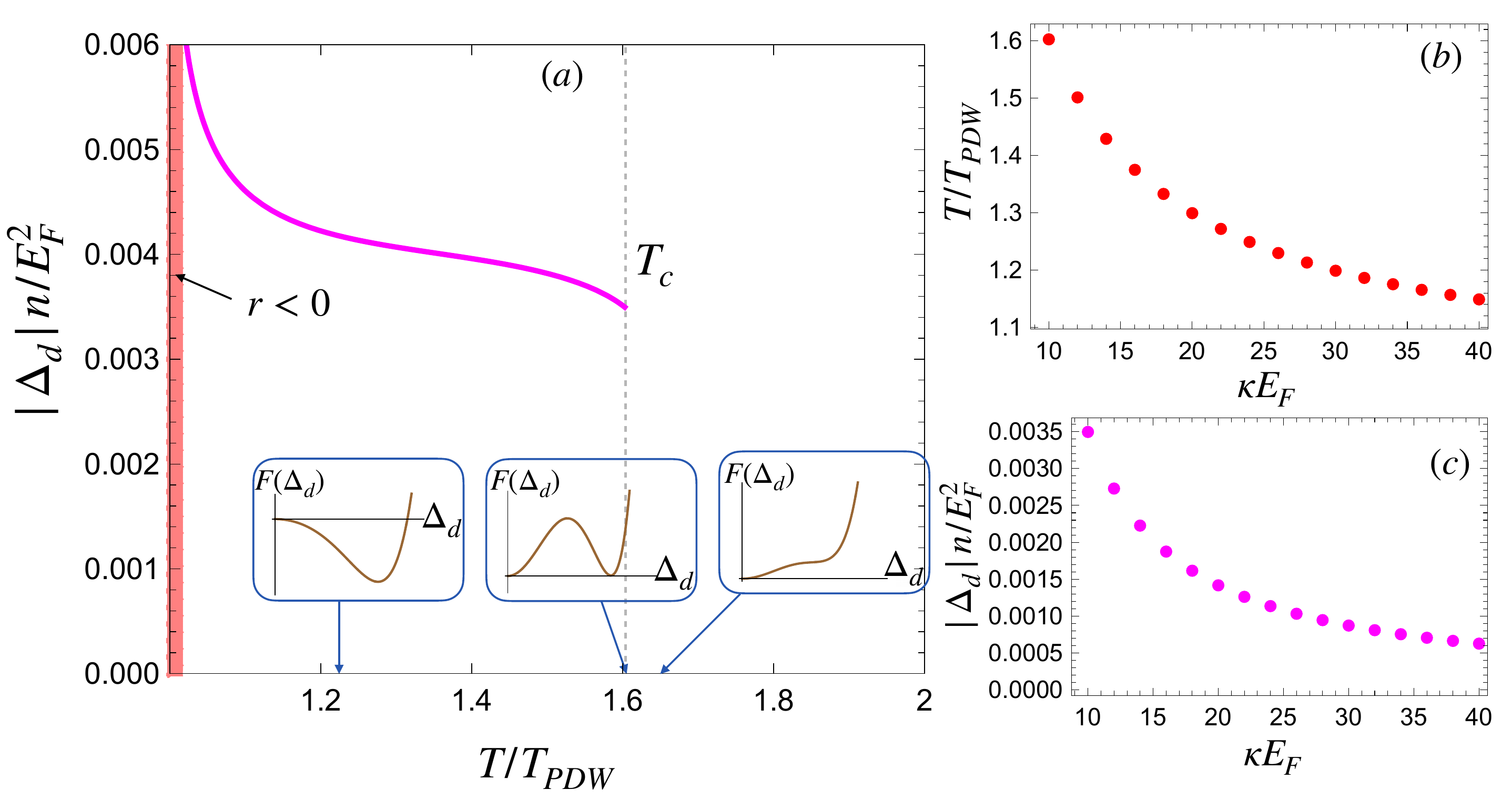}
  \caption{{Phase transition of the leading $d$-wave $4e$-SC.} (a) $4e$-SC gap amplitude as a function of $T$ obtained for  $T_{\text{PDW}}=0.1$ and $\kappa=10$. Insets: GL free energy as a function of $\Delta_d$. (b) and (c) The impact of $\kappa$ on $T_c$ and $\Delta_d$.}\label{fig:numerics}
\end{figure}

Complementary to the analytical approach when $\kappa T\gg 1$, we also solved numerically for saddle points using the full expressions for $A(T)$, $B(T)$, and $C(T)$ {in the Supplementary Note IV}~\cite{SM}, as is shown in Fig.~\ref{fig:numerics}.
We see that $|\Delta_d|$ increases monotonically as $T$ decreases below $T_c$. Once $|\Delta_d|$ becomes large enough, $r$ in Eq.~\eqref{eq:16} will eventually become negative, invalidating our perturbative analysis. In practice, in the region of $r<0$ one needs to keep expanding to higher orders of the PDW order parameters. We also show the free energy profile for $T>T_c$, $T=T_c$ and $T<T_c$, from which the nature of first order phase transition is seen directly. In order to show the impact of the stiffness $\kappa$ on various quantities for the $4e$-SC order, we show in Fig.~\ref{fig:numerics}{(b) and (c)} the plot of $T_c$ and $|\Delta_d(T_c)|$ as a function of $\kappa$. It is clear that a larger $\kappa$ yields a smaller $T_c$ and $|\Delta_d(T_c)|$. This is consistent with Eq.~\eqref{eq:tc} obtained in the limit that $\kappa T\gg 1$.



{It is possible that deep inside the $4e$-SC state, the translation symmetry is further broken, so that $\langle \rho_{\bp-\bq}\rangle \neq 0$. By symmetry, this is  when the primary PDW order develops, but the actual transition temperature for the PDW state does not coincide with the mean field $T_{\text{PDW}}$, and it can only set in via a Kosterlitz-Thouless transition (since a $U(1)$ translation symmetry is broken) at much lower temperature, which is beyond the scope of our GL analysis and calls for other approaches, e.g., numerics or the nonlinear sigma model in Ref.~\cite{Berg2009}.}

\noindent\textit{Outlook.---} 
The $d$-wave symmetry for the $4e$-SC order can be directly measured by phase-sensitive experiments, just like its $2e$-SC counterpart~\cite{Tsuei-2000}. Our theory has interesting implications for unconventional superconductors. There exist strong evidence for PDW in underdoped cuprates {such as in LBCO near 1/8 doping, where in-plane superconducting fluctuations, together with intertwined density-wave orders, develop at a higher temperature than the bulk superconducting temperature.~\cite{Agterberg} The PDW wave vector was proposed to be unidirectional within each Cu-O plane and alters between $x$ and $y$ ordering direction between neighboring planes \cite{PhysRevLett.99.127003}, which is consistent with recent measurements in $c$-axis transport~\cite{Shi2020}}. Our work points to a possibility of $4e$-SC with a relative sign change between neighboring Cu-O planes with perpendicular PDW wavevectors, although microscopic details, {such as whether Eq.~\eqref{eq:0} is approximately satisified}, call for a separate analysis. In addition, both PDW and $4e$-SC (and $6e$-SC) have been proposed to exist in the Kagome metal CsV$_3$Sb$_5$~\cite{yu2023nondegenerate,arxiv.2211.01388,Zhou2022,schwemmer2023pair,Scammell2023}.
It would be interesting to generalize our theory to hexagonal systems, which may be applied to CsV$_3$Sb$_5$.

{Finally, we note that our theoretical treatment of the $4e$-SC transition, like many previous analyses of vestigial orders~\cite{PhysRevB.85.024534,PhysRevB.90.035149,Fernandes2021,https://doi.org/10.48550/arxiv.2301.01344,hecker2023cascade}, is ultimately a mean-field theory, which focuses on certain channels of fluctuation while neglecting others. In our work, based on analytical calculations, we specifically neglected fluctuations in the nematic, CDW, and $s$-wave $4e$-SC channels. Nevertheless, it would be interesting to test our results via unbiased numerical simulations, such as quantum Monte Carlo. We leave this to future work.}

\begin{acknowledgments}
\noindent{\it Acknowledgments.---} The authors would like to thank Andrey Chubukov, Rafael Fernandes, Sri Raghu, Pavel Nosov, Hong Yao and Steven Kivelson for useful discussions. Y.-M. Wu acknowledges the Gordon and Betty Moore Foundation’s EPiQS Initiative through GBMF8686 for support at Stanford University. Y. Wang is supported by NSF under award number DMR-2045781.
  
 \noindent{\it Competing Interests.---}  The Authors declare no Competing Financial or Non-Financial Interests.
\end{acknowledgments}

 \noindent{\it Data availability.---} Data will be kept in UF database and will be available upon request.

 \noindent{\it Author contributions.---} 
Y.~Wu and Y.~Wang performed analytic calculations. Y.~Wu performed the numerical calculations. Both authors wrote the draft.

\bibliography{charge4e}

\begin{thebibliography}{63}%
\makeatletter
\providecommand \@ifxundefined [1]{%
 \@ifx{#1\undefined}
}%
\providecommand \@ifnum [1]{%
 \ifnum #1\expandafter \@firstoftwo
 \else \expandafter \@secondoftwo
 \fi
}%
\providecommand \@ifx [1]{%
 \ifx #1\expandafter \@firstoftwo
 \else \expandafter \@secondoftwo
 \fi
}%
\providecommand \natexlab [1]{#1}%
\providecommand \enquote  [1]{``#1''}%
\providecommand \bibnamefont  [1]{#1}%
\providecommand \bibfnamefont [1]{#1}%
\providecommand \citenamefont [1]{#1}%
\providecommand \href@noop [0]{\@secondoftwo}%
\providecommand \href [0]{\begingroup \@sanitize@url \@href}%
\providecommand \@href[1]{\@@startlink{#1}\@@href}%
\providecommand \@@href[1]{\endgroup#1\@@endlink}%
\providecommand \@sanitize@url [0]{\catcode `\\12\catcode `\$12\catcode
  `\&12\catcode `\#12\catcode `\^12\catcode `\_12\catcode `\%12\relax}%
\providecommand \@@startlink[1]{}%
\providecommand \@@endlink[0]{}%
\providecommand \url  [0]{\begingroup\@sanitize@url \@url }%
\providecommand \@url [1]{\endgroup\@href {#1}{\urlprefix }}%
\providecommand \urlprefix  [0]{URL }%
\providecommand \Eprint [0]{\href }%
\providecommand \doibase [0]{https://doi.org/}%
\providecommand \selectlanguage [0]{\@gobble}%
\providecommand \bibinfo  [0]{\@secondoftwo}%
\providecommand \bibfield  [0]{\@secondoftwo}%
\providecommand \translation [1]{[#1]}%
\providecommand \BibitemOpen [0]{}%
\providecommand \bibitemStop [0]{}%
\providecommand \bibitemNoStop [0]{.\EOS\space}%
\providecommand \EOS [0]{\spacefactor3000\relax}%
\providecommand \BibitemShut  [1]{\csname bibitem#1\endcsname}%
\let\auto@bib@innerbib\@empty
\bibitem [{\citenamefont {R\"opke}\ \emph {et~al.}(1998)\citenamefont
  {R\"opke}, \citenamefont {Schnell}, \citenamefont {Schuck},\ and\
  \citenamefont {Nozi\`eres}}]{PhysRevLett.80.3177}%
  \BibitemOpen
  \bibfield  {author} {\bibinfo {author} {\bibfnamefont {G.}~\bibnamefont
  {R\"opke}}, \bibinfo {author} {\bibfnamefont {A.}~\bibnamefont {Schnell}},
  \bibinfo {author} {\bibfnamefont {P.}~\bibnamefont {Schuck}},\ and\ \bibinfo
  {author} {\bibfnamefont {P.}~\bibnamefont {Nozi\`eres}},\ }\bibfield  {title}
  {\bibinfo {title} {Four-particle condensate in strongly coupled fermion
  systems},\ }\href {https://doi.org/10.1103/PhysRevLett.80.3177} {\bibfield
  {journal} {\bibinfo  {journal} {Phys. Rev. Lett.}\ }\textbf {\bibinfo
  {volume} {80}},\ \bibinfo {pages} {3177} (\bibinfo {year}
  {1998})}\BibitemShut {NoStop}%
\bibitem [{\citenamefont {Wu}(2005)}]{PhysRevLett.95.266404}%
  \BibitemOpen
  \bibfield  {author} {\bibinfo {author} {\bibfnamefont {C.}~\bibnamefont
  {Wu}},\ }\bibfield  {title} {\bibinfo {title} {Competing orders in
  one-dimensional spin-$3/2$ fermionic systems},\ }\href
  {https://doi.org/10.1103/PhysRevLett.95.266404} {\bibfield  {journal}
  {\bibinfo  {journal} {Phys. Rev. Lett.}\ }\textbf {\bibinfo {volume} {95}},\
  \bibinfo {pages} {266404} (\bibinfo {year} {2005})}\BibitemShut {NoStop}%
\bibitem [{\citenamefont {Aligia}\ \emph {et~al.}(2005)\citenamefont {Aligia},
  \citenamefont {Kampf},\ and\ \citenamefont
  {Mannhart}}]{PhysRevLett.94.247004}%
  \BibitemOpen
  \bibfield  {author} {\bibinfo {author} {\bibfnamefont {A.~A.}\ \bibnamefont
  {Aligia}}, \bibinfo {author} {\bibfnamefont {A.~P.}\ \bibnamefont {Kampf}},\
  and\ \bibinfo {author} {\bibfnamefont {J.}~\bibnamefont {Mannhart}},\
  }\bibfield  {title} {\bibinfo {title} {Quartet formation at $(100)/(110)$
  interfaces of $d$-wave superconductors},\ }\href
  {https://doi.org/10.1103/PhysRevLett.94.247004} {\bibfield  {journal}
  {\bibinfo  {journal} {Phys. Rev. Lett.}\ }\textbf {\bibinfo {volume} {94}},\
  \bibinfo {pages} {247004} (\bibinfo {year} {2005})}\BibitemShut {NoStop}%
\bibitem [{\citenamefont {Herland}\ \emph {et~al.}(2010)\citenamefont
  {Herland}, \citenamefont {Babaev},\ and\ \citenamefont
  {Sudb\o{}}}]{PhysRevB.82.134511}%
  \BibitemOpen
  \bibfield  {author} {\bibinfo {author} {\bibfnamefont {E.~V.}\ \bibnamefont
  {Herland}}, \bibinfo {author} {\bibfnamefont {E.}~\bibnamefont {Babaev}},\
  and\ \bibinfo {author} {\bibfnamefont {A.}~\bibnamefont {Sudb\o{}}},\
  }\bibfield  {title} {\bibinfo {title} {Phase transitions in a three
  dimensional $u(1)\ifmmode\times\else\texttimes\fi{}u(1)$ lattice london
  superconductor: Metallic superfluid and charge-$4e$ superconducting states},\
  }\href {https://doi.org/10.1103/PhysRevB.82.134511} {\bibfield  {journal}
  {\bibinfo  {journal} {Phys. Rev. B}\ }\textbf {\bibinfo {volume} {82}},\
  \bibinfo {pages} {134511} (\bibinfo {year} {2010})}\BibitemShut {NoStop}%
\bibitem [{\citenamefont {Svistunov}\ \emph {et~al.}(2015)\citenamefont
  {Svistunov}, \citenamefont {Babaev},\ and\ \citenamefont
  {Prokof'ev}}]{svistunov2015superfluid}%
  \BibitemOpen
  \bibfield  {author} {\bibinfo {author} {\bibfnamefont {B.}~\bibnamefont
  {Svistunov}}, \bibinfo {author} {\bibfnamefont {E.}~\bibnamefont {Babaev}},\
  and\ \bibinfo {author} {\bibfnamefont {N.}~\bibnamefont {Prokof'ev}},\ }\href
  {https://books.google.com/books?id=kW93CAAAQBAJ} {\emph {\bibinfo {title}
  {Superfluid States of Matter}}}\ (\bibinfo  {publisher} {CRC Press},\
  \bibinfo {year} {2015})\BibitemShut {NoStop}%
\bibitem [{\citenamefont {Li}\ \emph {et~al.}(2022)\citenamefont {Li},
  \citenamefont {Jiang},\ and\ \citenamefont {Hu}}]{li2022charge}%
  \BibitemOpen
  \bibfield  {author} {\bibinfo {author} {\bibfnamefont {P.}~\bibnamefont
  {Li}}, \bibinfo {author} {\bibfnamefont {K.}~\bibnamefont {Jiang}},\ and\
  \bibinfo {author} {\bibfnamefont {J.}~\bibnamefont {Hu}},\ }\href@noop {}
  {\bibinfo {title} {Charge 4$e$ superconductor: a wavefunction approach}}
  (\bibinfo {year} {2022}),\ \Eprint {https://arxiv.org/abs/2209.13905}
  {arXiv:2209.13905 [cond-mat.supr-con]} \BibitemShut {NoStop}%
\bibitem [{\citenamefont {Ge}\ \emph {et~al.}(2022)\citenamefont {Ge},
  \citenamefont {Wang}, \citenamefont {Xing}, \citenamefont {Yin},
  \citenamefont {Lei}, \citenamefont {Wang},\ and\ \citenamefont
  {Wang}}]{ge2022}%
  \BibitemOpen
  \bibfield  {author} {\bibinfo {author} {\bibfnamefont {J.}~\bibnamefont
  {Ge}}, \bibinfo {author} {\bibfnamefont {P.}~\bibnamefont {Wang}}, \bibinfo
  {author} {\bibfnamefont {Y.}~\bibnamefont {Xing}}, \bibinfo {author}
  {\bibfnamefont {Q.}~\bibnamefont {Yin}}, \bibinfo {author} {\bibfnamefont
  {H.}~\bibnamefont {Lei}}, \bibinfo {author} {\bibfnamefont {Z.}~\bibnamefont
  {Wang}},\ and\ \bibinfo {author} {\bibfnamefont {J.}~\bibnamefont {Wang}},\
  }\bibfield  {title} {\bibinfo {title} {Discovery of charge-4e and charge-6e
  superconductivity in kagome superconductor csv3sb5}\ }\href
  {https://doi.org/10.48550/ARXIV.2201.10352} {10.48550/ARXIV.2201.10352}
  (\bibinfo {year} {2022})\BibitemShut {NoStop}%
\bibitem [{\citenamefont {Gnezdilov}\ and\ \citenamefont
  {Wang}(2022)}]{Gnezdilov}%
  \BibitemOpen
  \bibfield  {author} {\bibinfo {author} {\bibfnamefont {N.~V.}\ \bibnamefont
  {Gnezdilov}}\ and\ \bibinfo {author} {\bibfnamefont {Y.}~\bibnamefont
  {Wang}},\ }\bibfield  {title} {\bibinfo {title} {Solvable model for a
  charge-$4e$ superconductor},\ }\href
  {https://doi.org/10.1103/PhysRevB.106.094508} {\bibfield  {journal} {\bibinfo
   {journal} {Phys. Rev. B}\ }\textbf {\bibinfo {volume} {106}},\ \bibinfo
  {pages} {094508} (\bibinfo {year} {2022})}\BibitemShut {NoStop}%
\bibitem [{\citenamefont {Jiang}\ \emph {et~al.}(2017)\citenamefont {Jiang},
  \citenamefont {Li}, \citenamefont {Kivelson},\ and\ \citenamefont
  {Yao}}]{Jiang2017}%
  \BibitemOpen
  \bibfield  {author} {\bibinfo {author} {\bibfnamefont {Y.-F.}\ \bibnamefont
  {Jiang}}, \bibinfo {author} {\bibfnamefont {Z.-X.}\ \bibnamefont {Li}},
  \bibinfo {author} {\bibfnamefont {S.~A.}\ \bibnamefont {Kivelson}},\ and\
  \bibinfo {author} {\bibfnamefont {H.}~\bibnamefont {Yao}},\ }\bibfield
  {title} {\bibinfo {title} {Charge-$4e$ superconductors: A majorana quantum
  monte carlo study},\ }\href {https://doi.org/10.1103/PhysRevB.95.241103}
  {\bibfield  {journal} {\bibinfo  {journal} {Phys. Rev. B}\ }\textbf {\bibinfo
  {volume} {95}},\ \bibinfo {pages} {241103} (\bibinfo {year}
  {2017})}\BibitemShut {NoStop}%
\bibitem [{\citenamefont {Volovik}(1992)}]{volovik1992exotic}%
  \BibitemOpen
  \bibfield  {author} {\bibinfo {author} {\bibfnamefont {G.}~\bibnamefont
  {Volovik}},\ }\href {https://books.google.com/books?id=gpvsCgAAQBAJ} {\emph
  {\bibinfo {title} {Exotic Properties Of Superfluid Helium 3}}},\ Series In
  Modern Condensed Matter Physics\ (\bibinfo  {publisher} {World Scientific
  Publishing Company},\ \bibinfo {year} {1992})\BibitemShut {NoStop}%
\bibitem [{\citenamefont {Moon}(2012)}]{PhysRevB.85.245123}%
  \BibitemOpen
  \bibfield  {author} {\bibinfo {author} {\bibfnamefont {E.-G.}\ \bibnamefont
  {Moon}},\ }\bibfield  {title} {\bibinfo {title} {Skyrmions with quadratic
  band touching fermions: A way to achieve charge $4e$ superconductivity},\
  }\href {https://doi.org/10.1103/PhysRevB.85.245123} {\bibfield  {journal}
  {\bibinfo  {journal} {Phys. Rev. B}\ }\textbf {\bibinfo {volume} {85}},\
  \bibinfo {pages} {245123} (\bibinfo {year} {2012})}\BibitemShut {NoStop}%
\bibitem [{\citenamefont {Fradkin}\ \emph {et~al.}(2015)\citenamefont
  {Fradkin}, \citenamefont {Kivelson},\ and\ \citenamefont
  {Tranquada}}]{RevModPhys.87.457}%
  \BibitemOpen
  \bibfield  {author} {\bibinfo {author} {\bibfnamefont {E.}~\bibnamefont
  {Fradkin}}, \bibinfo {author} {\bibfnamefont {S.~A.}\ \bibnamefont
  {Kivelson}},\ and\ \bibinfo {author} {\bibfnamefont {J.~M.}\ \bibnamefont
  {Tranquada}},\ }\bibfield  {title} {\bibinfo {title} {Colloquium: Theory of
  intertwined orders in high temperature superconductors},\ }\href
  {https://doi.org/10.1103/RevModPhys.87.457} {\bibfield  {journal} {\bibinfo
  {journal} {Rev. Mod. Phys.}\ }\textbf {\bibinfo {volume} {87}},\ \bibinfo
  {pages} {457} (\bibinfo {year} {2015})}\BibitemShut {NoStop}%
\bibitem [{\citenamefont {Babaev}\ \emph {et~al.}(2004)\citenamefont {Babaev},
  \citenamefont {Sudb{\o}},\ and\ \citenamefont {Ashcroft}}]{Babaev2004}%
  \BibitemOpen
  \bibfield  {author} {\bibinfo {author} {\bibfnamefont {E.}~\bibnamefont
  {Babaev}}, \bibinfo {author} {\bibfnamefont {A.}~\bibnamefont {Sudb{\o}}},\
  and\ \bibinfo {author} {\bibfnamefont {N.~W.}\ \bibnamefont {Ashcroft}},\
  }\bibfield  {title} {\bibinfo {title} {A superconductor to superfluid phase
  transition in liquid metallic hydrogen},\ }\href
  {https://doi.org/10.1038/nature02910} {\bibfield  {journal} {\bibinfo
  {journal} {Nature}\ }\textbf {\bibinfo {volume} {431}},\ \bibinfo {pages}
  {666} (\bibinfo {year} {2004})}\BibitemShut {NoStop}%
\bibitem [{\citenamefont {Agterberg}\ and\ \citenamefont
  {Tsunetsugu}(2008)}]{Agterberg2008}%
  \BibitemOpen
  \bibfield  {author} {\bibinfo {author} {\bibfnamefont {D.~F.}\ \bibnamefont
  {Agterberg}}\ and\ \bibinfo {author} {\bibfnamefont {H.}~\bibnamefont
  {Tsunetsugu}},\ }\bibfield  {title} {\bibinfo {title} {Dislocations and
  vortices in pair-density-wave superconductors},\ }\href
  {https://doi.org/10.1038/nphys999} {\bibfield  {journal} {\bibinfo  {journal}
  {Nature Physics}\ }\textbf {\bibinfo {volume} {4}},\ \bibinfo {pages} {639}
  (\bibinfo {year} {2008})}\BibitemShut {NoStop}%
\bibitem [{\citenamefont {Berg}\ \emph
  {et~al.}(2009{\natexlab{a}})\citenamefont {Berg}, \citenamefont {Fradkin},\
  and\ \citenamefont {Kivelson}}]{Berg2009}%
  \BibitemOpen
  \bibfield  {author} {\bibinfo {author} {\bibfnamefont {E.}~\bibnamefont
  {Berg}}, \bibinfo {author} {\bibfnamefont {E.}~\bibnamefont {Fradkin}},\ and\
  \bibinfo {author} {\bibfnamefont {S.~A.}\ \bibnamefont {Kivelson}},\
  }\bibfield  {title} {\bibinfo {title} {Charge-4e superconductivity from
  pair-density-wave order in certain high-temperature superconductors},\ }\href
  {https://doi.org/10.1038/nphys1389} {\bibfield  {journal} {\bibinfo
  {journal} {Nature Physics}\ }\textbf {\bibinfo {volume} {5}},\ \bibinfo
  {pages} {830} (\bibinfo {year} {2009}{\natexlab{a}})}\BibitemShut {NoStop}%
\bibitem [{\citenamefont {Radzihovsky}\ and\ \citenamefont
  {Vishwanath}(2009)}]{PhysRevLett.103.010404}%
  \BibitemOpen
  \bibfield  {author} {\bibinfo {author} {\bibfnamefont {L.}~\bibnamefont
  {Radzihovsky}}\ and\ \bibinfo {author} {\bibfnamefont {A.}~\bibnamefont
  {Vishwanath}},\ }\bibfield  {title} {\bibinfo {title} {Quantum liquid
  crystals in an imbalanced fermi gas: Fluctuations and fractional vortices in
  larkin-ovchinnikov states},\ }\href
  {https://doi.org/10.1103/PhysRevLett.103.010404} {\bibfield  {journal}
  {\bibinfo  {journal} {Phys. Rev. Lett.}\ }\textbf {\bibinfo {volume} {103}},\
  \bibinfo {pages} {010404} (\bibinfo {year} {2009})}\BibitemShut {NoStop}%
\bibitem [{\citenamefont {Agterberg}\ \emph {et~al.}(2011)\citenamefont
  {Agterberg}, \citenamefont {Geracie},\ and\ \citenamefont
  {Tsunetsugu}}]{Agterberg2011}%
  \BibitemOpen
  \bibfield  {author} {\bibinfo {author} {\bibfnamefont {D.~F.}\ \bibnamefont
  {Agterberg}}, \bibinfo {author} {\bibfnamefont {M.}~\bibnamefont {Geracie}},\
  and\ \bibinfo {author} {\bibfnamefont {H.}~\bibnamefont {Tsunetsugu}},\
  }\bibfield  {title} {\bibinfo {title} {Conventional and charge-six
  superfluids from melting hexagonal fulde-ferrell-larkin-ovchinnikov phases in
  two dimensions},\ }\href {https://doi.org/10.1103/PhysRevB.84.014513}
  {\bibfield  {journal} {\bibinfo  {journal} {Phys. Rev. B}\ }\textbf {\bibinfo
  {volume} {84}},\ \bibinfo {pages} {014513} (\bibinfo {year}
  {2011})}\BibitemShut {NoStop}%
\bibitem [{\citenamefont {Fernandes}\ and\ \citenamefont
  {Fu}(2021)}]{Fernandes2021}%
  \BibitemOpen
  \bibfield  {author} {\bibinfo {author} {\bibfnamefont {R.~M.}\ \bibnamefont
  {Fernandes}}\ and\ \bibinfo {author} {\bibfnamefont {L.}~\bibnamefont {Fu}},\
  }\bibfield  {title} {\bibinfo {title} {Charge-$4e$ superconductivity from
  multicomponent nematic pairing: Application to twisted bilayer graphene},\
  }\href {https://doi.org/10.1103/PhysRevLett.127.047001} {\bibfield  {journal}
  {\bibinfo  {journal} {Phys. Rev. Lett.}\ }\textbf {\bibinfo {volume} {127}},\
  \bibinfo {pages} {047001} (\bibinfo {year} {2021})}\BibitemShut {NoStop}%
\bibitem [{\citenamefont {Jian}\ \emph {et~al.}(2021)\citenamefont {Jian},
  \citenamefont {Huang},\ and\ \citenamefont {Yao}}]{Jian2021}%
  \BibitemOpen
  \bibfield  {author} {\bibinfo {author} {\bibfnamefont {S.-K.}\ \bibnamefont
  {Jian}}, \bibinfo {author} {\bibfnamefont {Y.}~\bibnamefont {Huang}},\ and\
  \bibinfo {author} {\bibfnamefont {H.}~\bibnamefont {Yao}},\ }\bibfield
  {title} {\bibinfo {title} {Charge-$4e$ superconductivity from nematic
  superconductors in two and three dimensions},\ }\href
  {https://doi.org/10.1103/PhysRevLett.127.227001} {\bibfield  {journal}
  {\bibinfo  {journal} {Phys. Rev. Lett.}\ }\textbf {\bibinfo {volume} {127}},\
  \bibinfo {pages} {227001} (\bibinfo {year} {2021})}\BibitemShut {NoStop}%
\bibitem [{\citenamefont {Liu}\ \emph {et~al.}(2023{\natexlab{a}})\citenamefont
  {Liu}, \citenamefont {Zhou}, \citenamefont {Wu},\ and\ \citenamefont
  {Yang}}]{liu2023charge}%
  \BibitemOpen
  \bibfield  {author} {\bibinfo {author} {\bibfnamefont {Y.-B.}\ \bibnamefont
  {Liu}}, \bibinfo {author} {\bibfnamefont {J.}~\bibnamefont {Zhou}}, \bibinfo
  {author} {\bibfnamefont {C.}~\bibnamefont {Wu}},\ and\ \bibinfo {author}
  {\bibfnamefont {F.}~\bibnamefont {Yang}},\ }\href@noop {} {\bibinfo {title}
  {Charge 4e superconductivity and chiral metal in the $45^\circ$-twisted
  bilayer cuprates and similar materials}} (\bibinfo {year}
  {2023}{\natexlab{a}}),\ \Eprint {https://arxiv.org/abs/2301.06357}
  {arXiv:2301.06357 [cond-mat.supr-con]} \BibitemShut {NoStop}%
\bibitem [{\citenamefont {Poduval}\ and\ \citenamefont
  {Scheurer}(2023)}]{https://doi.org/10.48550/arxiv.2301.01344}%
  \BibitemOpen
  \bibfield  {author} {\bibinfo {author} {\bibfnamefont {P.~P.}\ \bibnamefont
  {Poduval}}\ and\ \bibinfo {author} {\bibfnamefont {M.~S.}\ \bibnamefont
  {Scheurer}},\ }\href {https://doi.org/10.48550/ARXIV.2301.01344} {\bibinfo
  {title} {Vestigial singlet pairing in a fluctuating magnetic triplet
  superconductor: Applications to graphene moiré systems. arxiv.2301.01344}}
  (\bibinfo {year} {2023})\BibitemShut {NoStop}%
\bibitem [{\citenamefont {Hecker}\ \emph {et~al.}(2023)\citenamefont {Hecker},
  \citenamefont {Willa}, \citenamefont {Schmalian},\ and\ \citenamefont
  {Fernandes}}]{hecker2023cascade}%
  \BibitemOpen
  \bibfield  {author} {\bibinfo {author} {\bibfnamefont {M.}~\bibnamefont
  {Hecker}}, \bibinfo {author} {\bibfnamefont {R.}~\bibnamefont {Willa}},
  \bibinfo {author} {\bibfnamefont {J.}~\bibnamefont {Schmalian}},\ and\
  \bibinfo {author} {\bibfnamefont {R.~M.}\ \bibnamefont {Fernandes}},\
  }\href@noop {} {\bibinfo {title} {Cascade of vestigial orders in
  two-component superconductors: nematic, ferromagnetic, s-wave charge-4e, and
  d-wave charge-4e states. arxiv:2303.00653}} (\bibinfo {year} {2023}),\
  \Eprint {https://arxiv.org/abs/2303.00653} {arXiv:2303.00653
  [cond-mat.supr-con]} \BibitemShut {NoStop}%
\bibitem [{\citenamefont {Hecker}\ and\ \citenamefont
  {Fernandes}(2023)}]{hecker2023local}%
  \BibitemOpen
  \bibfield  {author} {\bibinfo {author} {\bibfnamefont {M.}~\bibnamefont
  {Hecker}}\ and\ \bibinfo {author} {\bibfnamefont {R.~M.}\ \bibnamefont
  {Fernandes}},\ }\href@noop {} {\bibinfo {title} {Local condensation of
  charge-$4e$ superconductivity at a nematic domain wall}} (\bibinfo {year}
  {2023}),\ \Eprint {https://arxiv.org/abs/2311.02005} {arXiv:2311.02005
  [cond-mat.supr-con]} \BibitemShut {NoStop}%
\bibitem [{\citenamefont {Liu}\ \emph {et~al.}(2023{\natexlab{b}})\citenamefont
  {Liu}, \citenamefont {Wang},\ and\ \citenamefont {Cui}}]{xiaoling2023}%
  \BibitemOpen
  \bibfield  {author} {\bibinfo {author} {\bibfnamefont {R.}~\bibnamefont
  {Liu}}, \bibinfo {author} {\bibfnamefont {W.}~\bibnamefont {Wang}},\ and\
  \bibinfo {author} {\bibfnamefont {X.}~\bibnamefont {Cui}},\ }\bibfield
  {title} {\bibinfo {title} {Quartet superfluid in two-dimensional
  mass-imbalanced fermi mixtures},\ }\href
  {https://doi.org/10.1103/PhysRevLett.131.193401} {\bibfield  {journal}
  {\bibinfo  {journal} {Phys. Rev. Lett.}\ }\textbf {\bibinfo {volume} {131}},\
  \bibinfo {pages} {193401} (\bibinfo {year} {2023}{\natexlab{b}})}\BibitemShut
  {NoStop}%
\bibitem [{\citenamefont {Agterberg}\ \emph {et~al.}(2020)\citenamefont
  {Agterberg}, \citenamefont {Davis}, \citenamefont {Edkins}, \citenamefont
  {Fradkin}, \citenamefont {Van~Harlingen}, \citenamefont {Kivelson},
  \citenamefont {Lee}, \citenamefont {Radzihovsky}, \citenamefont {Tranquada},\
  and\ \citenamefont {Wang}}]{Agterberg}%
  \BibitemOpen
  \bibfield  {author} {\bibinfo {author} {\bibfnamefont {D.~F.}\ \bibnamefont
  {Agterberg}}, \bibinfo {author} {\bibfnamefont {J.~S.}\ \bibnamefont
  {Davis}}, \bibinfo {author} {\bibfnamefont {S.~D.}\ \bibnamefont {Edkins}},
  \bibinfo {author} {\bibfnamefont {E.}~\bibnamefont {Fradkin}}, \bibinfo
  {author} {\bibfnamefont {D.~J.}\ \bibnamefont {Van~Harlingen}}, \bibinfo
  {author} {\bibfnamefont {S.~A.}\ \bibnamefont {Kivelson}}, \bibinfo {author}
  {\bibfnamefont {P.~A.}\ \bibnamefont {Lee}}, \bibinfo {author} {\bibfnamefont
  {L.}~\bibnamefont {Radzihovsky}}, \bibinfo {author} {\bibfnamefont {J.~M.}\
  \bibnamefont {Tranquada}},\ and\ \bibinfo {author} {\bibfnamefont
  {Y.}~\bibnamefont {Wang}},\ }\bibfield  {title} {\bibinfo {title} {The
  physics of pair-density waves: Cuprate superconductors and beyond},\ }\href
  {https://doi.org/10.1146/annurev-conmatphys-031119-050711} {\bibfield
  {journal} {\bibinfo  {journal} {Annual Review of Condensed Matter Physics}\
  }\textbf {\bibinfo {volume} {11}},\ \bibinfo {pages} {231} (\bibinfo {year}
  {2020})}\BibitemShut {NoStop}%
\bibitem [{\citenamefont {Fulde}\ and\ \citenamefont {Ferrell}(1964)}]{FF}%
  \BibitemOpen
  \bibfield  {author} {\bibinfo {author} {\bibfnamefont {P.}~\bibnamefont
  {Fulde}}\ and\ \bibinfo {author} {\bibfnamefont {R.~A.}\ \bibnamefont
  {Ferrell}},\ }\bibfield  {title} {\bibinfo {title} {Superconductivity in a
  strong spin-exchange field},\ }\href
  {https://doi.org/10.1103/PhysRev.135.A550} {\bibfield  {journal} {\bibinfo
  {journal} {Phys. Rev.}\ }\textbf {\bibinfo {volume} {135}},\ \bibinfo {pages}
  {A550} (\bibinfo {year} {1964})}\BibitemShut {NoStop}%
\bibitem [{\citenamefont {Larkin}\ and\ \citenamefont
  {Ovchinnikov}(1965)}]{LO}%
  \BibitemOpen
  \bibfield  {author} {\bibinfo {author} {\bibfnamefont {A.~I.}\ \bibnamefont
  {Larkin}}\ and\ \bibinfo {author} {\bibfnamefont {Y.~N.}\ \bibnamefont
  {Ovchinnikov}},\ }\bibfield  {title} {\bibinfo {title} {Nonuniform state of
  superconductors},\ }\href {https://www.osti.gov/biblio/4653415} {\bibfield
  {journal} {\bibinfo  {journal} {Sov. Phys. JETP}\ }\textbf {\bibinfo {volume}
  {20}},\ \bibinfo {pages} {762} (\bibinfo {year} {1965})}\BibitemShut
  {NoStop}%
\bibitem [{\citenamefont {Agosta}(2018)}]{cryst8070285}%
  \BibitemOpen
  \bibfield  {author} {\bibinfo {author} {\bibfnamefont {C.~C.}\ \bibnamefont
  {Agosta}},\ }\bibfield  {title} {\bibinfo {title} {Inhomogeneous
  superconductivity in organic and related superconductors},\ }\href
  {https://doi.org/10.3390/cryst8070285} {\bibfield  {journal} {\bibinfo
  {journal} {Crystals}\ }\textbf {\bibinfo {volume} {8}},\ \bibinfo {pages}
  {285} (\bibinfo {year} {2018})}\BibitemShut {NoStop}%
\bibitem [{\citenamefont {Matsuda}\ and\ \citenamefont
  {Shimahara}(2007)}]{Matsuda2007}%
  \BibitemOpen
  \bibfield  {author} {\bibinfo {author} {\bibfnamefont {Y.}~\bibnamefont
  {Matsuda}}\ and\ \bibinfo {author} {\bibfnamefont {H.}~\bibnamefont
  {Shimahara}},\ }\bibfield  {title} {\bibinfo {title}
  {Fulde–ferrell–larkin–ovchinnikov state in heavy fermion
  superconductors},\ }\href {https://doi.org/10.1143/JPSJ.76.051005} {\bibfield
   {journal} {\bibinfo  {journal} {Journal of the Physical Society of Japan}\
  }\textbf {\bibinfo {volume} {76}},\ \bibinfo {pages} {051005} (\bibinfo
  {year} {2007})}\BibitemShut {NoStop}%
\bibitem [{\citenamefont {Gurevich}(2010)}]{Gurevich}%
  \BibitemOpen
  \bibfield  {author} {\bibinfo {author} {\bibfnamefont {A.}~\bibnamefont
  {Gurevich}},\ }\bibfield  {title} {\bibinfo {title} {Upper critical field and
  the fulde-ferrel-larkin-ovchinnikov transition in multiband
  superconductors},\ }\href {https://doi.org/10.1103/PhysRevB.82.184504}
  {\bibfield  {journal} {\bibinfo  {journal} {Phys. Rev. B}\ }\textbf {\bibinfo
  {volume} {82}},\ \bibinfo {pages} {184504} (\bibinfo {year}
  {2010})}\BibitemShut {NoStop}%
\bibitem [{\citenamefont {Cho}\ \emph {et~al.}(2017)\citenamefont {Cho},
  \citenamefont {Yang}, \citenamefont {Yuan}, \citenamefont {Shen},
  \citenamefont {Wolf},\ and\ \citenamefont {Lortz}}]{FeSC}%
  \BibitemOpen
  \bibfield  {author} {\bibinfo {author} {\bibfnamefont {C.-w.}\ \bibnamefont
  {Cho}}, \bibinfo {author} {\bibfnamefont {J.~H.}\ \bibnamefont {Yang}},
  \bibinfo {author} {\bibfnamefont {N.~F.~Q.}\ \bibnamefont {Yuan}}, \bibinfo
  {author} {\bibfnamefont {J.}~\bibnamefont {Shen}}, \bibinfo {author}
  {\bibfnamefont {T.}~\bibnamefont {Wolf}},\ and\ \bibinfo {author}
  {\bibfnamefont {R.}~\bibnamefont {Lortz}},\ }\bibfield  {title} {\bibinfo
  {title} {Thermodynamic evidence for the fulde-ferrell-larkin-ovchinnikov
  state in the ${\mathrm{kfe}}_{2}{\mathrm{as}}_{2}$ superconductor},\ }\href
  {https://doi.org/10.1103/PhysRevLett.119.217002} {\bibfield  {journal}
  {\bibinfo  {journal} {Phys. Rev. Lett.}\ }\textbf {\bibinfo {volume} {119}},\
  \bibinfo {pages} {217002} (\bibinfo {year} {2017})}\BibitemShut {NoStop}%
\bibitem [{\citenamefont {Liu}\ \emph {et~al.}(2021)\citenamefont {Liu},
  \citenamefont {Chong}, \citenamefont {Sharma},\ and\ \citenamefont
  {Davis}}]{doi:10.1126/science.abd4607}%
  \BibitemOpen
  \bibfield  {author} {\bibinfo {author} {\bibfnamefont {X.}~\bibnamefont
  {Liu}}, \bibinfo {author} {\bibfnamefont {Y.~X.}\ \bibnamefont {Chong}},
  \bibinfo {author} {\bibfnamefont {R.}~\bibnamefont {Sharma}},\ and\ \bibinfo
  {author} {\bibfnamefont {J.~C.~S.}\ \bibnamefont {Davis}},\ }\bibfield
  {title} {\bibinfo {title} {Discovery of a cooper-pair density wave state in a
  transition-metal dichalcogenide},\ }\href
  {https://doi.org/10.1126/science.abd4607} {\bibfield  {journal} {\bibinfo
  {journal} {Science}\ }\textbf {\bibinfo {volume} {372}},\ \bibinfo {pages}
  {1447} (\bibinfo {year} {2021})}\BibitemShut {NoStop}%
\bibitem [{\citenamefont {Wu}\ \emph {et~al.}(2023{\natexlab{a}})\citenamefont
  {Wu}, \citenamefont {Wu},\ and\ \citenamefont
  {Yao}}]{PhysRevLett.130.126001}%
  \BibitemOpen
  \bibfield  {author} {\bibinfo {author} {\bibfnamefont {Y.-M.}\ \bibnamefont
  {Wu}}, \bibinfo {author} {\bibfnamefont {Z.}~\bibnamefont {Wu}},\ and\
  \bibinfo {author} {\bibfnamefont {H.}~\bibnamefont {Yao}},\ }\bibfield
  {title} {\bibinfo {title} {Pair-density-wave and chiral superconductivity in
  twisted bilayer transition metal dichalcogenides},\ }\href
  {https://doi.org/10.1103/PhysRevLett.130.126001} {\bibfield  {journal}
  {\bibinfo  {journal} {Phys. Rev. Lett.}\ }\textbf {\bibinfo {volume} {130}},\
  \bibinfo {pages} {126001} (\bibinfo {year} {2023}{\natexlab{a}})}\BibitemShut
  {NoStop}%
\bibitem [{\citenamefont {Wu}\ \emph {et~al.}(2023{\natexlab{b}})\citenamefont
  {Wu}, \citenamefont {Wu},\ and\ \citenamefont {Wu}}]{PhysRevB.107.045122}%
  \BibitemOpen
  \bibfield  {author} {\bibinfo {author} {\bibfnamefont {Z.}~\bibnamefont
  {Wu}}, \bibinfo {author} {\bibfnamefont {Y.-M.}\ \bibnamefont {Wu}},\ and\
  \bibinfo {author} {\bibfnamefont {F.}~\bibnamefont {Wu}},\ }\bibfield
  {title} {\bibinfo {title} {Pair density wave and loop current promoted by van
  hove singularities in moir\'e systems},\ }\href
  {https://doi.org/10.1103/PhysRevB.107.045122} {\bibfield  {journal} {\bibinfo
   {journal} {Phys. Rev. B}\ }\textbf {\bibinfo {volume} {107}},\ \bibinfo
  {pages} {045122} (\bibinfo {year} {2023}{\natexlab{b}})}\BibitemShut
  {NoStop}%
\bibitem [{\citenamefont {Cho}\ \emph {et~al.}(2012)\citenamefont {Cho},
  \citenamefont {Bardarson}, \citenamefont {Lu},\ and\ \citenamefont
  {Moore}}]{PhysRevB.86.214514}%
  \BibitemOpen
  \bibfield  {author} {\bibinfo {author} {\bibfnamefont {G.~Y.}\ \bibnamefont
  {Cho}}, \bibinfo {author} {\bibfnamefont {J.~H.}\ \bibnamefont {Bardarson}},
  \bibinfo {author} {\bibfnamefont {Y.-M.}\ \bibnamefont {Lu}},\ and\ \bibinfo
  {author} {\bibfnamefont {J.~E.}\ \bibnamefont {Moore}},\ }\bibfield  {title}
  {\bibinfo {title} {Superconductivity of doped weyl semimetals:
  Finite-momentum pairing and electronic analog of the ${}^{3}$he-$a$ phase},\
  }\href {https://doi.org/10.1103/PhysRevB.86.214514} {\bibfield  {journal}
  {\bibinfo  {journal} {Phys. Rev. B}\ }\textbf {\bibinfo {volume} {86}},\
  \bibinfo {pages} {214514} (\bibinfo {year} {2012})}\BibitemShut {NoStop}%
\bibitem [{\citenamefont {Bednik}\ \emph {et~al.}(2015)\citenamefont {Bednik},
  \citenamefont {Zyuzin},\ and\ \citenamefont {Burkov}}]{PhysRevB.92.035153}%
  \BibitemOpen
  \bibfield  {author} {\bibinfo {author} {\bibfnamefont {G.}~\bibnamefont
  {Bednik}}, \bibinfo {author} {\bibfnamefont {A.~A.}\ \bibnamefont {Zyuzin}},\
  and\ \bibinfo {author} {\bibfnamefont {A.~A.}\ \bibnamefont {Burkov}},\
  }\bibfield  {title} {\bibinfo {title} {Superconductivity in weyl metals},\
  }\href {https://doi.org/10.1103/PhysRevB.92.035153} {\bibfield  {journal}
  {\bibinfo  {journal} {Phys. Rev. B}\ }\textbf {\bibinfo {volume} {92}},\
  \bibinfo {pages} {035153} (\bibinfo {year} {2015})}\BibitemShut {NoStop}%
\bibitem [{\citenamefont {Shaffer}\ \emph {et~al.}(2022)\citenamefont
  {Shaffer}, \citenamefont {Burnell},\ and\ \citenamefont
  {Fernandes}}]{https://doi.org/10.48550/arxiv.2209.14469}%
  \BibitemOpen
  \bibfield  {author} {\bibinfo {author} {\bibfnamefont {D.}~\bibnamefont
  {Shaffer}}, \bibinfo {author} {\bibfnamefont {F.~J.}\ \bibnamefont
  {Burnell}},\ and\ \bibinfo {author} {\bibfnamefont {R.~M.}\ \bibnamefont
  {Fernandes}},\ }\href {https://doi.org/10.48550/ARXIV.2209.14469} {\bibinfo
  {title} {Weak-coupling theory of pair density-wave instabilities in
  transition metal dichalcogenides. arxiv.2209.14469}} (\bibinfo {year}
  {2022})\BibitemShut {NoStop}%
\bibitem [{\citenamefont {Shaffer}\ and\ \citenamefont
  {Santos}(2022)}]{https://doi.org/10.48550/arxiv.2210.16324}%
  \BibitemOpen
  \bibfield  {author} {\bibinfo {author} {\bibfnamefont {D.}~\bibnamefont
  {Shaffer}}\ and\ \bibinfo {author} {\bibfnamefont {L.~H.}\ \bibnamefont
  {Santos}},\ }\href {https://doi.org/10.48550/ARXIV.2210.16324} {\bibinfo
  {title} {Triplet pair-density wave superconductivity on the $\pi$-flux square
  lattice. arxiv.2210.16324}} (\bibinfo {year} {2022})\BibitemShut {NoStop}%
\bibitem [{\citenamefont {Shaffer}\ \emph {et~al.}(2021)\citenamefont
  {Shaffer}, \citenamefont {Wang},\ and\ \citenamefont
  {Santos}}]{PhysRevB.104.184501}%
  \BibitemOpen
  \bibfield  {author} {\bibinfo {author} {\bibfnamefont {D.}~\bibnamefont
  {Shaffer}}, \bibinfo {author} {\bibfnamefont {J.}~\bibnamefont {Wang}},\ and\
  \bibinfo {author} {\bibfnamefont {L.~H.}\ \bibnamefont {Santos}},\ }\bibfield
   {title} {\bibinfo {title} {Theory of hofstadter superconductors},\ }\href
  {https://doi.org/10.1103/PhysRevB.104.184501} {\bibfield  {journal} {\bibinfo
   {journal} {Phys. Rev. B}\ }\textbf {\bibinfo {volume} {104}},\ \bibinfo
  {pages} {184501} (\bibinfo {year} {2021})}\BibitemShut {NoStop}%
\bibitem [{\citenamefont {Berg}\ \emph
  {et~al.}(2009{\natexlab{b}})\citenamefont {Berg}, \citenamefont {Fradkin},
  \citenamefont {Kivelson},\ and\ \citenamefont {Tranquada}}]{Berg_2009}%
  \BibitemOpen
  \bibfield  {author} {\bibinfo {author} {\bibfnamefont {E.}~\bibnamefont
  {Berg}}, \bibinfo {author} {\bibfnamefont {E.}~\bibnamefont {Fradkin}},
  \bibinfo {author} {\bibfnamefont {S.~A.}\ \bibnamefont {Kivelson}},\ and\
  \bibinfo {author} {\bibfnamefont {J.~M.}\ \bibnamefont {Tranquada}},\
  }\bibfield  {title} {\bibinfo {title} {Striped superconductors: how spin,
  charge and superconducting orders intertwine in the cuprates},\ }\href
  {https://doi.org/10.1088/1367-2630/11/11/115004} {\bibfield  {journal}
  {\bibinfo  {journal} {New Journal of Physics}\ }\textbf {\bibinfo {volume}
  {11}},\ \bibinfo {pages} {115004} (\bibinfo {year}
  {2009}{\natexlab{b}})}\BibitemShut {NoStop}%
\bibitem [{\citenamefont {H\"ucker}\ \emph {et~al.}(2011)\citenamefont
  {H\"ucker}, \citenamefont {v.~Zimmermann}, \citenamefont {Gu}, \citenamefont
  {Xu}, \citenamefont {Wen}, \citenamefont {Xu}, \citenamefont {Kang},
  \citenamefont {Zheludev},\ and\ \citenamefont {Tranquada}}]{Zimmermann2011}%
  \BibitemOpen
  \bibfield  {author} {\bibinfo {author} {\bibfnamefont {M.}~\bibnamefont
  {H\"ucker}}, \bibinfo {author} {\bibfnamefont {M.}~\bibnamefont
  {v.~Zimmermann}}, \bibinfo {author} {\bibfnamefont {G.~D.}\ \bibnamefont
  {Gu}}, \bibinfo {author} {\bibfnamefont {Z.~J.}\ \bibnamefont {Xu}}, \bibinfo
  {author} {\bibfnamefont {J.~S.}\ \bibnamefont {Wen}}, \bibinfo {author}
  {\bibfnamefont {G.}~\bibnamefont {Xu}}, \bibinfo {author} {\bibfnamefont
  {H.~J.}\ \bibnamefont {Kang}}, \bibinfo {author} {\bibfnamefont
  {A.}~\bibnamefont {Zheludev}},\ and\ \bibinfo {author} {\bibfnamefont
  {J.~M.}\ \bibnamefont {Tranquada}},\ }\bibfield  {title} {\bibinfo {title}
  {Stripe order in superconducting
  la${}_{2\ensuremath{-}x}$ba${}_{x}$cuo${}_{4}$
  ($0.095\ensuremath{\leqslant}x\ensuremath{\leqslant}0.155$)},\ }\href
  {https://doi.org/10.1103/PhysRevB.83.104506} {\bibfield  {journal} {\bibinfo
  {journal} {Phys. Rev. B}\ }\textbf {\bibinfo {volume} {83}},\ \bibinfo
  {pages} {104506} (\bibinfo {year} {2011})}\BibitemShut {NoStop}%
\bibitem [{\citenamefont {Wang}\ \emph
  {et~al.}(2015{\natexlab{a}})\citenamefont {Wang}, \citenamefont {Agterberg},\
  and\ \citenamefont {Chubukov}}]{PhysRevLett.114.197001}%
  \BibitemOpen
  \bibfield  {author} {\bibinfo {author} {\bibfnamefont {Y.}~\bibnamefont
  {Wang}}, \bibinfo {author} {\bibfnamefont {D.~F.}\ \bibnamefont
  {Agterberg}},\ and\ \bibinfo {author} {\bibfnamefont {A.}~\bibnamefont
  {Chubukov}},\ }\bibfield  {title} {\bibinfo {title} {Coexistence of
  charge-density-wave and pair-density-wave orders in underdoped cuprates},\
  }\href {https://doi.org/10.1103/PhysRevLett.114.197001} {\bibfield  {journal}
  {\bibinfo  {journal} {Phys. Rev. Lett.}\ }\textbf {\bibinfo {volume} {114}},\
  \bibinfo {pages} {197001} (\bibinfo {year} {2015}{\natexlab{a}})}\BibitemShut
  {NoStop}%
\bibitem [{\citenamefont {Berg}\ \emph {et~al.}(2007)\citenamefont {Berg},
  \citenamefont {Fradkin}, \citenamefont {Kim}, \citenamefont {Kivelson},
  \citenamefont {Oganesyan}, \citenamefont {Tranquada},\ and\ \citenamefont
  {Zhang}}]{PhysRevLett.99.127003}%
  \BibitemOpen
  \bibfield  {author} {\bibinfo {author} {\bibfnamefont {E.}~\bibnamefont
  {Berg}}, \bibinfo {author} {\bibfnamefont {E.}~\bibnamefont {Fradkin}},
  \bibinfo {author} {\bibfnamefont {E.-A.}\ \bibnamefont {Kim}}, \bibinfo
  {author} {\bibfnamefont {S.~A.}\ \bibnamefont {Kivelson}}, \bibinfo {author}
  {\bibfnamefont {V.}~\bibnamefont {Oganesyan}}, \bibinfo {author}
  {\bibfnamefont {J.~M.}\ \bibnamefont {Tranquada}},\ and\ \bibinfo {author}
  {\bibfnamefont {S.~C.}\ \bibnamefont {Zhang}},\ }\bibfield  {title} {\bibinfo
  {title} {Dynamical layer decoupling in a stripe-ordered high-${T}_{c}$
  superconductor},\ }\href {https://doi.org/10.1103/PhysRevLett.99.127003}
  {\bibfield  {journal} {\bibinfo  {journal} {Phys. Rev. Lett.}\ }\textbf
  {\bibinfo {volume} {99}},\ \bibinfo {pages} {127003} (\bibinfo {year}
  {2007})}\BibitemShut {NoStop}%
\bibitem [{\citenamefont {Wang}\ \emph
  {et~al.}(2015{\natexlab{b}})\citenamefont {Wang}, \citenamefont {Agterberg},\
  and\ \citenamefont {Chubukov}}]{PhysRevB.91.115103}%
  \BibitemOpen
  \bibfield  {author} {\bibinfo {author} {\bibfnamefont {Y.}~\bibnamefont
  {Wang}}, \bibinfo {author} {\bibfnamefont {D.~F.}\ \bibnamefont
  {Agterberg}},\ and\ \bibinfo {author} {\bibfnamefont {A.}~\bibnamefont
  {Chubukov}},\ }\bibfield  {title} {\bibinfo {title} {Interplay between pair-
  and charge-density-wave orders in underdoped cuprates},\ }\href
  {https://doi.org/10.1103/PhysRevB.91.115103} {\bibfield  {journal} {\bibinfo
  {journal} {Phys. Rev. B}\ }\textbf {\bibinfo {volume} {91}},\ \bibinfo
  {pages} {115103} (\bibinfo {year} {2015}{\natexlab{b}})}\BibitemShut
  {NoStop}%
\bibitem [{\citenamefont {Lee}(2014)}]{PALee2014}%
  \BibitemOpen
  \bibfield  {author} {\bibinfo {author} {\bibfnamefont {P.~A.}\ \bibnamefont
  {Lee}},\ }\bibfield  {title} {\bibinfo {title} {Amperean pairing and the
  pseudogap phase of cuprate superconductors},\ }\href
  {https://doi.org/10.1103/PhysRevX.4.031017} {\bibfield  {journal} {\bibinfo
  {journal} {Phys. Rev. X}\ }\textbf {\bibinfo {volume} {4}},\ \bibinfo {pages}
  {031017} (\bibinfo {year} {2014})}\BibitemShut {NoStop}%
\bibitem [{\citenamefont {Nie}\ \emph {et~al.}(2014)\citenamefont {Nie},
  \citenamefont {Tarjus},\ and\ \citenamefont {Kivelson}}]{Nie2014}%
  \BibitemOpen
  \bibfield  {author} {\bibinfo {author} {\bibfnamefont {L.}~\bibnamefont
  {Nie}}, \bibinfo {author} {\bibfnamefont {G.}~\bibnamefont {Tarjus}},\ and\
  \bibinfo {author} {\bibfnamefont {S.~A.}\ \bibnamefont {Kivelson}},\
  }\bibfield  {title} {\bibinfo {title} {Quenched disorder and vestigial
  nematicity in the pseudogap regime of the cuprates},\ }\href
  {https://doi.org/10.1073/pnas.1406019111} {\bibfield  {journal} {\bibinfo
  {journal} {Proceedings of the National Academy of Sciences}\ }\textbf
  {\bibinfo {volume} {111}},\ \bibinfo {pages} {7980} (\bibinfo {year}
  {2014})}\BibitemShut {NoStop}%
\bibitem [{\citenamefont {Setty}\ \emph {et~al.}(2021)\citenamefont {Setty},
  \citenamefont {Fanfarillo},\ and\ \citenamefont
  {Hirschfeld}}]{setty2021microscopic}%
  \BibitemOpen
  \bibfield  {author} {\bibinfo {author} {\bibfnamefont {C.}~\bibnamefont
  {Setty}}, \bibinfo {author} {\bibfnamefont {L.}~\bibnamefont {Fanfarillo}},\
  and\ \bibinfo {author} {\bibfnamefont {P.~J.}\ \bibnamefont {Hirschfeld}},\
  }\href@noop {} {\bibinfo {title} {Microscopic mechanism for fluctuating pair
  density wave}} (\bibinfo {year} {2021}),\ \Eprint
  {https://arxiv.org/abs/2110.13138} {arXiv:2110.13138 [cond-mat.supr-con]}
  \BibitemShut {NoStop}%
\bibitem [{\citenamefont {Setty}\ \emph {et~al.}(2022)\citenamefont {Setty},
  \citenamefont {Zhao}, \citenamefont {Fanfarillo}, \citenamefont {Huang},
  \citenamefont {Hirschfeld}, \citenamefont {Phillips},\ and\ \citenamefont
  {Yang}}]{setty2022exact}%
  \BibitemOpen
  \bibfield  {author} {\bibinfo {author} {\bibfnamefont {C.}~\bibnamefont
  {Setty}}, \bibinfo {author} {\bibfnamefont {J.}~\bibnamefont {Zhao}},
  \bibinfo {author} {\bibfnamefont {L.}~\bibnamefont {Fanfarillo}}, \bibinfo
  {author} {\bibfnamefont {E.~W.}\ \bibnamefont {Huang}}, \bibinfo {author}
  {\bibfnamefont {P.~J.}\ \bibnamefont {Hirschfeld}}, \bibinfo {author}
  {\bibfnamefont {P.~W.}\ \bibnamefont {Phillips}},\ and\ \bibinfo {author}
  {\bibfnamefont {K.}~\bibnamefont {Yang}},\ }\href@noop {} {\bibinfo {title}
  {Exact solution for finite center-of-mass momentum cooper pairing}} (\bibinfo
  {year} {2022}),\ \Eprint {https://arxiv.org/abs/2209.10568} {arXiv:2209.10568
  [cond-mat.supr-con]} \BibitemShut {NoStop}%
\bibitem [{\citenamefont {Wu}\ \emph {et~al.}(2023{\natexlab{c}})\citenamefont
  {Wu}, \citenamefont {Nosov}, \citenamefont {Patel},\ and\ \citenamefont
  {Raghu}}]{PhysRevLett.130.026001}%
  \BibitemOpen
  \bibfield  {author} {\bibinfo {author} {\bibfnamefont {Y.-M.}\ \bibnamefont
  {Wu}}, \bibinfo {author} {\bibfnamefont {P.~A.}\ \bibnamefont {Nosov}},
  \bibinfo {author} {\bibfnamefont {A.~A.}\ \bibnamefont {Patel}},\ and\
  \bibinfo {author} {\bibfnamefont {S.}~\bibnamefont {Raghu}},\ }\bibfield
  {title} {\bibinfo {title} {Pair density wave order from electron repulsion},\
  }\href {https://doi.org/10.1103/PhysRevLett.130.026001} {\bibfield  {journal}
  {\bibinfo  {journal} {Phys. Rev. Lett.}\ }\textbf {\bibinfo {volume} {130}},\
  \bibinfo {pages} {026001} (\bibinfo {year} {2023}{\natexlab{c}})}\BibitemShut
  {NoStop}%
\bibitem [{\citenamefont {Huang}\ \emph {et~al.}(2022)\citenamefont {Huang},
  \citenamefont {Han}, \citenamefont {Kivelson},\ and\ \citenamefont
  {Yao}}]{Huang2022}%
  \BibitemOpen
  \bibfield  {author} {\bibinfo {author} {\bibfnamefont {K.~S.}\ \bibnamefont
  {Huang}}, \bibinfo {author} {\bibfnamefont {Z.}~\bibnamefont {Han}}, \bibinfo
  {author} {\bibfnamefont {S.~A.}\ \bibnamefont {Kivelson}},\ and\ \bibinfo
  {author} {\bibfnamefont {H.}~\bibnamefont {Yao}},\ }\bibfield  {title}
  {\bibinfo {title} {Pair-density-wave in the strong coupling limit of the
  holstein-hubbard model},\ }\href {https://doi.org/10.1038/s41535-022-00426-w}
  {\bibfield  {journal} {\bibinfo  {journal} {npj Quantum Materials}\ }\textbf
  {\bibinfo {volume} {7}},\ \bibinfo {pages} {17} (\bibinfo {year}
  {2022})}\BibitemShut {NoStop}%
\bibitem [{\citenamefont {Jiang}(2021)}]{Jiang2021}%
  \BibitemOpen
  \bibfield  {author} {\bibinfo {author} {\bibfnamefont {H.-C.}\ \bibnamefont
  {Jiang}},\ }\bibfield  {title} {\bibinfo {title} {Superconductivity in the
  doped quantum spin liquid on the triangular lattice},\ }\href
  {https://doi.org/10.1038/s41535-021-00375-w} {\bibfield  {journal} {\bibinfo
  {journal} {npj Quantum Materials}\ }\textbf {\bibinfo {volume} {6}},\
  \bibinfo {pages} {71} (\bibinfo {year} {2021})}\BibitemShut {NoStop}%
\bibitem [{\citenamefont {Li}\ \emph {et~al.}(2023)\citenamefont {Li},
  \citenamefont {Li}, \citenamefont {Wang}, \citenamefont {Wan}, \citenamefont
  {Yang},\ and\ \citenamefont {Wen}}]{Li2023}%
  \BibitemOpen
  \bibfield  {author} {\bibinfo {author} {\bibfnamefont {H.}~\bibnamefont
  {Li}}, \bibinfo {author} {\bibfnamefont {H.}~\bibnamefont {Li}}, \bibinfo
  {author} {\bibfnamefont {Z.}~\bibnamefont {Wang}}, \bibinfo {author}
  {\bibfnamefont {S.}~\bibnamefont {Wan}}, \bibinfo {author} {\bibfnamefont
  {H.}~\bibnamefont {Yang}},\ and\ \bibinfo {author} {\bibfnamefont {H.-H.}\
  \bibnamefont {Wen}},\ }\bibfield  {title} {\bibinfo {title} {Low-energy gap
  emerging from confined nematic states in extremely underdoped cuprate
  superconductors},\ }\href {https://doi.org/10.1038/s41535-023-00552-z}
  {\bibfield  {journal} {\bibinfo  {journal} {npj Quantum Materials}\ }\textbf
  {\bibinfo {volume} {8}},\ \bibinfo {pages} {18} (\bibinfo {year}
  {2023})}\BibitemShut {NoStop}%
\bibitem [{\citenamefont {Peng}\ \emph {et~al.}(2021)\citenamefont {Peng},
  \citenamefont {Jiang}, \citenamefont {Devereaux},\ and\ \citenamefont
  {Jiang}}]{Peng2021}%
  \BibitemOpen
  \bibfield  {author} {\bibinfo {author} {\bibfnamefont {C.}~\bibnamefont
  {Peng}}, \bibinfo {author} {\bibfnamefont {Y.-F.}\ \bibnamefont {Jiang}},
  \bibinfo {author} {\bibfnamefont {T.~P.}\ \bibnamefont {Devereaux}},\ and\
  \bibinfo {author} {\bibfnamefont {H.-C.}\ \bibnamefont {Jiang}},\ }\bibfield
  {title} {\bibinfo {title} {Precursor of pair-density wave in doping kitaev
  spin liquid on the honeycomb lattice},\ }\href
  {https://doi.org/10.1038/s41535-021-00363-0} {\bibfield  {journal} {\bibinfo
  {journal} {npj Quantum Materials}\ }\textbf {\bibinfo {volume} {6}},\
  \bibinfo {pages} {64} (\bibinfo {year} {2021})}\BibitemShut {NoStop}%
\bibitem [{SM()}]{SM}%
  \BibitemOpen
  \href@noop {} {\bibinfo {title} {See the supplementary material for
  information about i) why the models that we use are free from fine tuning,
  ii) how to obtain the complete expansion of the pdw fields to eighth order,
  iii) how the vestigial cdw order is suppressed iv) how to obtain the gl free
  energy for the $d$-wave charge-$4e$ superconductivity and v) demonstrate the
  equivalence between hubbard-stratonavich transformation and the use of
  lagrangian multipliers.}}\BibitemShut {Stop}%
\bibitem [{\citenamefont {Wang}\ and\ \citenamefont
  {Chubukov}(2014)}]{PhysRevB.90.035149}%
  \BibitemOpen
  \bibfield  {author} {\bibinfo {author} {\bibfnamefont {Y.}~\bibnamefont
  {Wang}}\ and\ \bibinfo {author} {\bibfnamefont {A.}~\bibnamefont
  {Chubukov}},\ }\bibfield  {title} {\bibinfo {title} {Charge-density-wave
  order with momentum $(2q,0)$ and $(0,2q)$ within the spin-fermion model:
  Continuous and discrete symmetry breaking, preemptive composite order, and
  relation to pseudogap in hole-doped cuprates},\ }\href
  {https://doi.org/10.1103/PhysRevB.90.035149} {\bibfield  {journal} {\bibinfo
  {journal} {Phys. Rev. B}\ }\textbf {\bibinfo {volume} {90}},\ \bibinfo
  {pages} {035149} (\bibinfo {year} {2014})}\BibitemShut {NoStop}%
\bibitem [{\citenamefont {Fernandes}\ \emph {et~al.}(2012)\citenamefont
  {Fernandes}, \citenamefont {Chubukov}, \citenamefont {Knolle}, \citenamefont
  {Eremin},\ and\ \citenamefont {Schmalian}}]{PhysRevB.85.024534}%
  \BibitemOpen
  \bibfield  {author} {\bibinfo {author} {\bibfnamefont {R.~M.}\ \bibnamefont
  {Fernandes}}, \bibinfo {author} {\bibfnamefont {A.~V.}\ \bibnamefont
  {Chubukov}}, \bibinfo {author} {\bibfnamefont {J.}~\bibnamefont {Knolle}},
  \bibinfo {author} {\bibfnamefont {I.}~\bibnamefont {Eremin}},\ and\ \bibinfo
  {author} {\bibfnamefont {J.}~\bibnamefont {Schmalian}},\ }\bibfield  {title}
  {\bibinfo {title} {Preemptive nematic order, pseudogap, and orbital order in
  the iron pnictides},\ }\href {https://doi.org/10.1103/PhysRevB.85.024534}
  {\bibfield  {journal} {\bibinfo  {journal} {Phys. Rev. B}\ }\textbf {\bibinfo
  {volume} {85}},\ \bibinfo {pages} {024534} (\bibinfo {year}
  {2012})}\BibitemShut {NoStop}%
\bibitem [{\citenamefont {Tsuei}\ and\ \citenamefont
  {Kirtley}(2000)}]{Tsuei-2000}%
  \BibitemOpen
  \bibfield  {author} {\bibinfo {author} {\bibfnamefont {C.~C.}\ \bibnamefont
  {Tsuei}}\ and\ \bibinfo {author} {\bibfnamefont {J.~R.}\ \bibnamefont
  {Kirtley}},\ }\bibfield  {title} {\bibinfo {title} {Pairing symmetry in
  cuprate superconductors},\ }\href {https://doi.org/10.1103/RevModPhys.72.969}
  {\bibfield  {journal} {\bibinfo  {journal} {Rev. Mod. Phys.}\ }\textbf
  {\bibinfo {volume} {72}},\ \bibinfo {pages} {969} (\bibinfo {year}
  {2000})}\BibitemShut {NoStop}%
\bibitem [{\citenamefont {Shi}\ \emph {et~al.}(2020)\citenamefont {Shi},
  \citenamefont {Baity}, \citenamefont {Terzic}, \citenamefont {Sasagawa},\
  and\ \citenamefont {Popovi{\'{c}}}}]{Shi2020}%
  \BibitemOpen
  \bibfield  {author} {\bibinfo {author} {\bibfnamefont {Z.}~\bibnamefont
  {Shi}}, \bibinfo {author} {\bibfnamefont {P.~G.}\ \bibnamefont {Baity}},
  \bibinfo {author} {\bibfnamefont {J.}~\bibnamefont {Terzic}}, \bibinfo
  {author} {\bibfnamefont {T.}~\bibnamefont {Sasagawa}},\ and\ \bibinfo
  {author} {\bibfnamefont {D.}~\bibnamefont {Popovi{\'{c}}}},\ }\bibfield
  {title} {\bibinfo {title} {Pair density wave at high magnetic fields in
  cuprates with charge and spin orders},\ }\href
  {https://doi.org/10.1038/s41467-020-17138-z} {\bibfield  {journal} {\bibinfo
  {journal} {Nature Communications}\ }\textbf {\bibinfo {volume} {11}},\
  \bibinfo {pages} {3323} (\bibinfo {year} {2020})}\BibitemShut {NoStop}%
\bibitem [{\citenamefont {Yu}(2023)}]{yu2023nondegenerate}%
  \BibitemOpen
  \bibfield  {author} {\bibinfo {author} {\bibfnamefont {Y.}~\bibnamefont
  {Yu}},\ }\href@noop {} {\bibinfo {title} {Non-degenerate surface pair density
  wave in the kagome superconductor csv$_3$sb$_5$ -- application to vestigial
  orders}} (\bibinfo {year} {2023}),\ \Eprint
  {https://arxiv.org/abs/2210.00023} {arXiv:2210.00023 [cond-mat.supr-con]}
  \BibitemShut {NoStop}%
\bibitem [{\citenamefont {Wu}\ \emph {et~al.}(2022)\citenamefont {Wu},
  \citenamefont {Thomale},\ and\ \citenamefont {Raghu}}]{arxiv.2211.01388}%
  \BibitemOpen
  \bibfield  {author} {\bibinfo {author} {\bibfnamefont {Y.-M.}\ \bibnamefont
  {Wu}}, \bibinfo {author} {\bibfnamefont {R.}~\bibnamefont {Thomale}},\ and\
  \bibinfo {author} {\bibfnamefont {S.}~\bibnamefont {Raghu}},\ }\href
  {https://doi.org/10.48550/ARXIV.2211.01388} {\bibinfo {title} {Sublattice
  interference promotes pair density wave order in kagome metals.
  arxiv.2211.01388}} (\bibinfo {year} {2022})\BibitemShut {NoStop}%
\bibitem [{\citenamefont {Zhou}\ and\ \citenamefont {Wang}(2022)}]{Zhou2022}%
  \BibitemOpen
  \bibfield  {author} {\bibinfo {author} {\bibfnamefont {S.}~\bibnamefont
  {Zhou}}\ and\ \bibinfo {author} {\bibfnamefont {Z.}~\bibnamefont {Wang}},\
  }\bibfield  {title} {\bibinfo {title} {Chern fermi pocket, topological pair
  density wave, and charge-4e and charge-6e superconductivity in kagom{\'e}
  superconductors},\ }\href {https://doi.org/10.1038/s41467-022-34832-2}
  {\bibfield  {journal} {\bibinfo  {journal} {Nature Communications}\ }\textbf
  {\bibinfo {volume} {13}},\ \bibinfo {pages} {7288} (\bibinfo {year}
  {2022})}\BibitemShut {NoStop}%
\bibitem [{\citenamefont {Schwemmer}\ \emph {et~al.}(2023)\citenamefont
  {Schwemmer}, \citenamefont {Hohmann}, \citenamefont {Dürrnagel},
  \citenamefont {Potten}, \citenamefont {Beyer}, \citenamefont {Rachel},
  \citenamefont {Wu}, \citenamefont {Raghu}, \citenamefont {Müller},
  \citenamefont {Hanke},\ and\ \citenamefont {Thomale}}]{schwemmer2023pair}%
  \BibitemOpen
  \bibfield  {author} {\bibinfo {author} {\bibfnamefont {T.}~\bibnamefont
  {Schwemmer}}, \bibinfo {author} {\bibfnamefont {H.}~\bibnamefont {Hohmann}},
  \bibinfo {author} {\bibfnamefont {M.}~\bibnamefont {Dürrnagel}}, \bibinfo
  {author} {\bibfnamefont {J.}~\bibnamefont {Potten}}, \bibinfo {author}
  {\bibfnamefont {J.}~\bibnamefont {Beyer}}, \bibinfo {author} {\bibfnamefont
  {S.}~\bibnamefont {Rachel}}, \bibinfo {author} {\bibfnamefont {Y.-M.}\
  \bibnamefont {Wu}}, \bibinfo {author} {\bibfnamefont {S.}~\bibnamefont
  {Raghu}}, \bibinfo {author} {\bibfnamefont {T.}~\bibnamefont {Müller}},
  \bibinfo {author} {\bibfnamefont {W.}~\bibnamefont {Hanke}},\ and\ \bibinfo
  {author} {\bibfnamefont {R.}~\bibnamefont {Thomale}},\ }\href@noop {}
  {\bibinfo {title} {Pair density wave instability in the kagome hubbard
  model}} (\bibinfo {year} {2023}),\ \Eprint {https://arxiv.org/abs/2302.08517}
  {arXiv:2302.08517 [cond-mat.str-el]} \BibitemShut {NoStop}%
\bibitem [{\citenamefont {Scammell}\ \emph {et~al.}(2023)\citenamefont
  {Scammell}, \citenamefont {Ingham}, \citenamefont {Li},\ and\ \citenamefont
  {Sushkov}}]{Scammell2023}%
  \BibitemOpen
  \bibfield  {author} {\bibinfo {author} {\bibfnamefont {H.~D.}\ \bibnamefont
  {Scammell}}, \bibinfo {author} {\bibfnamefont {J.}~\bibnamefont {Ingham}},
  \bibinfo {author} {\bibfnamefont {T.}~\bibnamefont {Li}},\ and\ \bibinfo
  {author} {\bibfnamefont {O.~P.}\ \bibnamefont {Sushkov}},\ }\bibfield
  {title} {\bibinfo {title} {Chiral excitonic order from twofold van hove
  singularities in kagome metals},\ }\href
  {https://doi.org/10.1038/s41467-023-35987-2} {\bibfield  {journal} {\bibinfo
  {journal} {Nature Communications}\ }\textbf {\bibinfo {volume} {14}},\
  \bibinfo {pages} {605} (\bibinfo {year} {2023})}\BibitemShut {NoStop}%
\end{thebibliography}%


\begin{thebibliography}{4}%
\makeatletter
\providecommand \@ifxundefined [1]{%
 \@ifx{#1\undefined}
}%
\providecommand \@ifnum [1]{%
 \ifnum #1\expandafter \@firstoftwo
 \else \expandafter \@secondoftwo
 \fi
}%
\providecommand \@ifx [1]{%
 \ifx #1\expandafter \@firstoftwo
 \else \expandafter \@secondoftwo
 \fi
}%
\providecommand \natexlab [1]{#1}%
\providecommand \enquote  [1]{``#1''}%
\providecommand \bibnamefont  [1]{#1}%
\providecommand \bibfnamefont [1]{#1}%
\providecommand \citenamefont [1]{#1}%
\providecommand \href@noop [0]{\@secondoftwo}%
\providecommand \href [0]{\begingroup \@sanitize@url \@href}%
\providecommand \@href[1]{\@@startlink{#1}\@@href}%
\providecommand \@@href[1]{\endgroup#1\@@endlink}%
\providecommand \@sanitize@url [0]{\catcode `\\12\catcode `\$12\catcode
  `\&12\catcode `\#12\catcode `\^12\catcode `\_12\catcode `\%12\relax}%
\providecommand \@@startlink[1]{}%
\providecommand \@@endlink[0]{}%
\providecommand \url  [0]{\begingroup\@sanitize@url \@url }%
\providecommand \@url [1]{\endgroup\@href {#1}{\urlprefix }}%
\providecommand \urlprefix  [0]{URL }%
\providecommand \Eprint [0]{\href }%
\providecommand \doibase [0]{https://doi.org/}%
\providecommand \selectlanguage [0]{\@gobble}%
\providecommand \bibinfo  [0]{\@secondoftwo}%
\providecommand \bibfield  [0]{\@secondoftwo}%
\providecommand \translation [1]{[#1]}%
\providecommand \BibitemOpen [0]{}%
\providecommand \bibitemStop [0]{}%
\providecommand \bibitemNoStop [0]{.\EOS\space}%
\providecommand \EOS [0]{\spacefactor3000\relax}%
\providecommand \BibitemShut  [1]{\csname bibitem#1\endcsname}%
\let\auto@bib@innerbib\@empty
\bibitem [{\citenamefont {Wu}\ \emph {et~al.}(2023)\citenamefont {Wu},
  \citenamefont {Nosov}, \citenamefont {Patel},\ and\ \citenamefont
  {Raghu}}]{PhysRevLett.130.026001}%
  \BibitemOpen
  \bibfield  {author} {\bibinfo {author} {\bibfnamefont {Y.-M.}\ \bibnamefont
  {Wu}}, \bibinfo {author} {\bibfnamefont {P.~A.}\ \bibnamefont {Nosov}},
  \bibinfo {author} {\bibfnamefont {A.~A.}\ \bibnamefont {Patel}},\ and\
  \bibinfo {author} {\bibfnamefont {S.}~\bibnamefont {Raghu}},\ }\bibfield
  {title} {\bibinfo {title} {Pair density wave order from electron repulsion},\
  }\href {https://doi.org/10.1103/PhysRevLett.130.026001} {\bibfield  {journal}
  {\bibinfo  {journal} {Phys. Rev. Lett.}\ }\textbf {\bibinfo {volume} {130}},\
  \bibinfo {pages} {026001} (\bibinfo {year} {2023})}\BibitemShut {NoStop}%
\bibitem [{\citenamefont {Wang}\ \emph {et~al.}(2015)\citenamefont {Wang},
  \citenamefont {Agterberg},\ and\ \citenamefont
  {Chubukov}}]{PhysRevLett.114.197001}%
  \BibitemOpen
  \bibfield  {author} {\bibinfo {author} {\bibfnamefont {Y.}~\bibnamefont
  {Wang}}, \bibinfo {author} {\bibfnamefont {D.~F.}\ \bibnamefont
  {Agterberg}},\ and\ \bibinfo {author} {\bibfnamefont {A.}~\bibnamefont
  {Chubukov}},\ }\bibfield  {title} {\bibinfo {title} {Coexistence of
  charge-density-wave and pair-density-wave orders in underdoped cuprates},\
  }\href {https://doi.org/10.1103/PhysRevLett.114.197001} {\bibfield  {journal}
  {\bibinfo  {journal} {Phys. Rev. Lett.}\ }\textbf {\bibinfo {volume} {114}},\
  \bibinfo {pages} {197001} (\bibinfo {year} {2015})}\BibitemShut {NoStop}%
\bibitem [{Note1()}]{Note1}%
  \BibitemOpen
  \bibinfo {note} {$\lambda _d$ and $\lambda _d^*$, as well as $\Delta _d$ and
  $\Delta _d^*$, are independent fields. The integration domains for $\lambda
  _d$, $\lambda _d^*$, and $\mu $ are along the imaginary axis, but their
  saddle points may not be.}\BibitemShut {Stop}%
\bibitem [{\citenamefont {Wang}\ and\ \citenamefont
  {Chubukov}(2014)}]{PhysRevB.90.035149}%
  \BibitemOpen
  \bibfield  {author} {\bibinfo {author} {\bibfnamefont {Y.}~\bibnamefont
  {Wang}}\ and\ \bibinfo {author} {\bibfnamefont {A.}~\bibnamefont
  {Chubukov}},\ }\bibfield  {title} {\bibinfo {title} {Charge-density-wave
  order with momentum $(2q,0)$ and $(0,2q)$ within the spin-fermion model:
  Continuous and discrete symmetry breaking, preemptive composite order, and
  relation to pseudogap in hole-doped cuprates},\ }\href
  {https://doi.org/10.1103/PhysRevB.90.035149} {\bibfield  {journal} {\bibinfo
  {journal} {Phys. Rev. B}\ }\textbf {\bibinfo {volume} {90}},\ \bibinfo
  {pages} {035149} (\bibinfo {year} {2014})}\BibitemShut {NoStop}%
\end{thebibliography}%

\end{document}


\title{Supplementary Material for ``$d$-wave Charge-$4e$ Superconductivity From Fluctuating  Pair Density Waves''}
\author{Yi-Ming Wu}
\affiliation{Stanford Institute for Theoretical Physics, Stanford
  University, Stanford, California 94305, USA}
\author{Yuxuan Wang}
\affiliation{Department of Physics, University of Florida, Gainesville, Florida 32611, USA}

\date{\today}

\maketitle
\setcounter{equation}{0}
\setcounter{figure}{0}
\setcounter{table}{0}
\renewcommand{\theequation}{S\arabic{equation}}
\renewcommand{\thefigure}{Supplementary Figure \arabic{figure}}
\renewcommand{\thesection}{SUPPLEMENTARY NOTE \Roman{section}}
\renewcommand{\figurename}{}


\section{The fine-tuning-free models} 
\label{sec:the_fine_tuning_free_model}
In the main text, we have made the statement that in a four-fold rotational system with four different PDW momenta, the charge-4e superconductivity channel is enhanced as long as the the PDW momentum magnitude $|\bm{Q}|$ and the Fermi momentum $\bm{k}_F$ satisfies a specific relation
\begin{equation}
  |\bm{Q}|=\sqrt{2}|\bm{k}_F\mod(\pi,0)|.\label{eq:condition}
\end{equation}
At a first glance, one might think this relation requires some fine tuning of the Fermi surfaces. However, it does not.
To see this, we consider the two situations discussed in Fig.~1(c) and (d) in the main text. 

\begin{figure}
   \includegraphics[width=12cm]{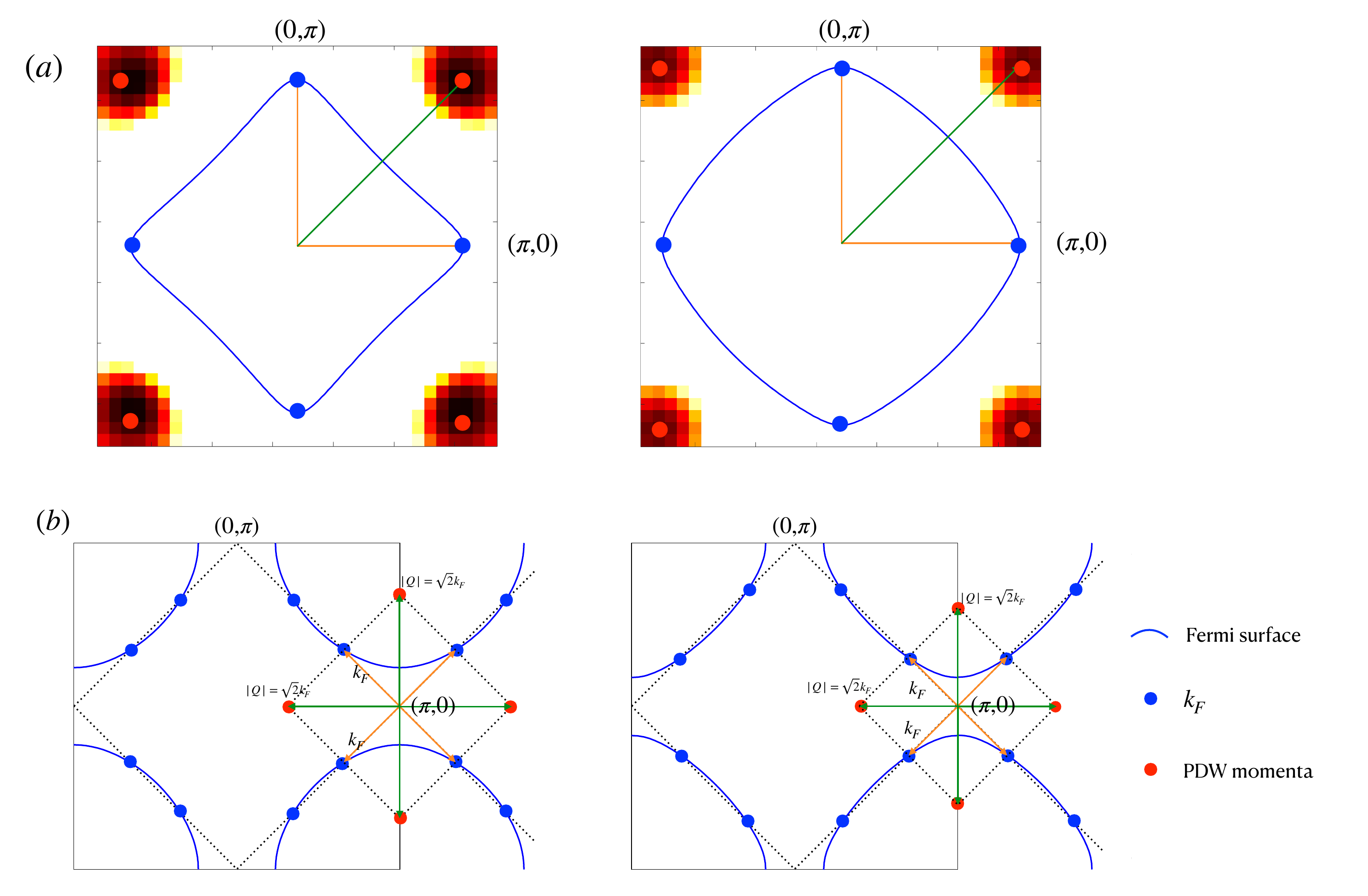}
   \caption{Illustration of the condition in Eq.~\eqref{eq:condition} for various situations in two different models. In both cases, changing electron (hole) filling, hopping amplitudes and other parameters will result in Fermi surfaces of different shapes, but the condition in Eq.~\eqref{eq:condition} can be easily satisfied without fine tuning.}\label{fig:noFT}
 \end{figure}

In the first model, the main contribution to the PDW pairing comes from those fermions close to the Van Hove singularity points at $(\pi,0)$ or $(0,\pi)$, since these places are where the fermion density of states maximizes. In \ref{fig:noFT}(a) we show two typical examples of this model. The Fermi surfaces {(blue curves)} are obtained by a tight binding model on a square lattice with either positive or negative next nearest neighbor hopping, and the onset of PDW has been demonstrated in Ref.~\cite{PhysRevLett.130.026001}.  {Reproducing the calculation there, we find that the PDW wave vectors are close to $(\pi, \pi)$. In \ref{fig:noFT}(a),  the heat map shows the inverse of the PDW susceptibility near the ordering vectors}.  We see that 
as long as the four-fold rotational symmetry is respected {and the system is close to Van Hove filling}, i.e. the four blue points {near the Van Hove points} form a square, then the PDW momenta (red points) automatically satisfies Eq.~\eqref{eq:condition}, regardless of the filling, and hopping parameters of the model. Therefore, it is free of fine tuning. 

Another example is given by the hot spot theory of superconductivity on a square lattice, which is demonstrate in \ref{fig:noFT} (b). Here we also choose two generic sets of parameters and obtain two configurations. In this model, the blue points, which are the intersections between the Fermi surface and the umklapp surfaces, are called hot spots. These points are the most likely positions where fermions can be scattered by the collective anti-ferromagnetic fluctuations carrying momentum $(\pi,\pi)$, and the pairing mechanism is driven by this collective fluctuation mediated interactions~\cite{PhysRevLett.114.197001}. Again, as long as the four-fold rotation symmetry is respected, the four closest hot spots form a square, and the resulting PDW momentum automatically satisfies Eq.~\eqref{eq:condition}. Note here both $\bm{k}_F$ and $\bm{Q}$ are measured from either $(\pi,0)$ or $(0,\pi)$.

From the above discussion and \ref{fig:noFT} we see that changing the electron (hole) filing fraction, or the hopping amplitude, or adding longer range hoppings, will surely change the shape of the Fermi surfaces. But as long as the four-fold rotation symmetry is present, the four active fermion spots (blue points) form a square, which is inscribed in another square formed from the resulting PDW momenta (red points). The condition in Eq.~\eqref{eq:condition} is the consequence of this simple geometric relation between the two squares, and thus is free from fine tuning. 

\section{Details of finding leading contributions up to $\mathcal{O}(|\Psi|^8)$} 
\label{sec:details_of_constructing_the_gl_theory}

\begin{figure}[t]
  \includegraphics[width=6cm]{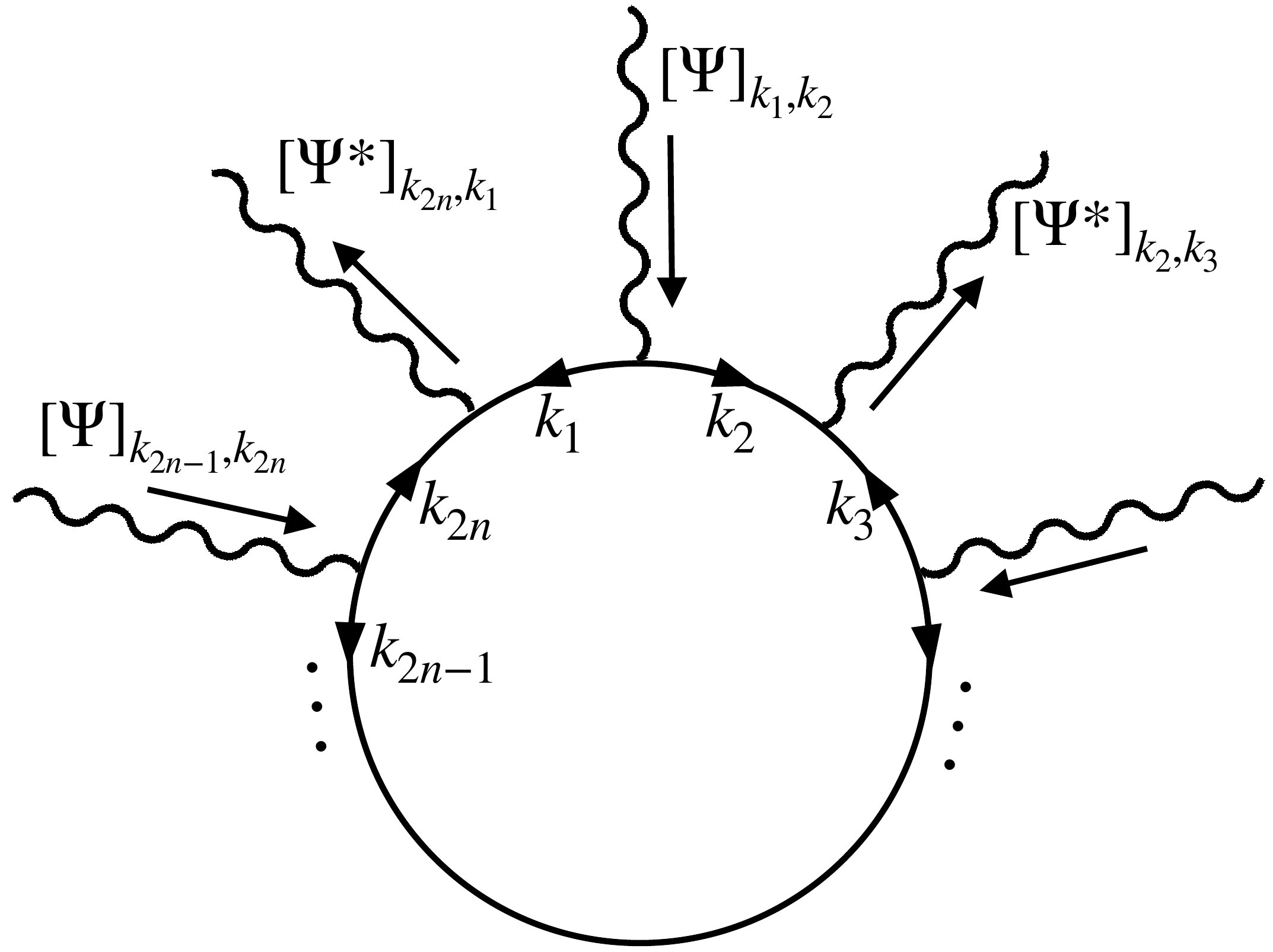}
  \caption{A general diagrammatic representation for higher-order interactions of PDW bosons.}\label{fig:diagramN}
\end{figure}

In obtaining the Ginzburg-Landau theory for the primary PDW orders, we integrate out the fermions and obtain 
\begin{equation}
  -\text{tr}\ln \mathcal{G}^{-1}\label{eq:act1}
\end{equation}
in the action. The matrix $\mathcal{G}$ is defined as  $\mathcal{G}^{-1}=G_0^{-1}+\hat\Delta$ and 
\begin{equation}
G_0^{-1}=
  \begin{pmatrix}
    G_p^{-1} & 0\\
    0 & G_h^{-1}
  \end{pmatrix}, ~~~
  \hat\Delta=
  \begin{pmatrix}
    0 & \Psi\\
    \Psi^\dagger& 0
  \end{pmatrix}\label{eq:mat}
\end{equation}
The particle and hole Green's function $G_p$ and $G_h$ are diagonal in frequency-momentum space. For instance, $[G_p]_{k,k'}=G_p(k)\delta_{k,k'}$. While $\Psi$ and $\Psi^\dagger$ is not diagonal in momentum space. In particular, we have $[\Psi]_{k,k'}=\sum_{i}\Psi_{\bQ_i}\delta_{k,\bQ_i-k'}$ and $[\Psi^\dagger]_{k,k'}=\sum_{i}\Psi^\dagger_{\bQ_i}\delta_{k,-\bQ_i-k'}$ with $\bQ_i\in\{\pm\bQ,\pm\bP\}$. Expanding Eq.~\eqref{eq:act1} in terms of $\Psi_{\bQ_i}$, we obtain
\begin{equation}
  \begin{aligned}
    -\text{tr}\ln \mathcal{G}^{-1}&=-\text{tr}\ln G_0^{-1}-\text{tr}\ln(1+G_0\hat\Delta)\\
    &=-\text{tr}\ln G_0^{-1}+\sum_{n=1}^\infty\frac{1}{2n}\text{tr}(G_0\hat\Delta)^{2n}.
  \end{aligned}
\end{equation}
Given the matrix structure in Eq.~\eqref{eq:mat}, it is easy to see that $G_0$  can be obtained by simply taking the inverse of the diagonal elements of $G_0^{-1}$. The $n$-th element in the series is given by
\begin{equation}
  \begin{aligned}
    \frac{1}{2n}\sum_{\{k_i\}}&[G_p]_{k_1}[\Psi]_{k_1,k_2}[G_h]_{k_2}[\Psi^*]_{k_2,k_3}...\\
    &...[G_p]_{k_{2n-1}}[\Psi]_{k_{2n-1},k_{2n}}[G_h]_{k_{2n}}[\Psi^*]_{k_{2n},k_1},\label{eq:expandG}
  \end{aligned}
\end{equation}
which is also shown diagrammatically in \ref{fig:diagramN}.

In our special situation, all the Green's functions should be $G_i, i=1,...,4$, which are
\begin{equation}
  \begin{aligned}
    &G_1(k)=\frac{1}{i\omega_n-v_Fk_x},G_2(k)=\frac{-1}{i\omega_n-v_Fk_y},\\
    &G_3(k)=\frac{1}{i\omega_n+v_Fk_x},G_4(k)=\frac{-1}{i\omega_n+v_Fk_y}.\\
  \end{aligned}\label{eq:invG}
\end{equation}
where $G_1$ and $G_3$ are particle Green's functions, while $G_2$ and $G_4$ are hole Green's functions. 
Given a set of Green's functions, the external momentum can be determined by momentum conservation. Therefore, each diagram uniquely corresponds to one permutation of the Green's function set.  We can develop a simple algorithm to numerate all the possible combinations for any $n$ based on Eq.~\eqref{eq:expandG}. The algorithm processes as follows,
\begin{enumerate}
  \item For a given $n$, specify all possible set $S$ for $2n$ Green's functions that satisfy the momentum conservation rule. For instance, for $n=2$, one possible set is $S=\{1,2,3,4\}$, meaning all the leading contributions corresponds to all the permutations of this set. 
  \item Generate all possible permutations for the set $S$, and keep only those nonequivalent ones. 
  \item Identify external legs for each nonequivalent permutations based on momentum conservation. 
  \item collect all the terms from different permutations for a given set, repeat the procedure for all sets. 
\end{enumerate}

It is easy to implement this algorithm to numerate all possible terms that satisfy the above conditions. 
For the quartic term (4-leg diagrams), we find the following contributions,
\begin{equation}
  \beta_1|\Psi_i|^4+\beta_2|\Psi_{\pm\bP}|^2|\Psi_{\pm\bQ}|^2+\beta(\Psi_{\bP}\Psi_{-\bP}\Psi^*_{\bQ}\Psi^*_{-\bQ}+h.c.).
\end{equation}
Note the last $\beta$-term is represented by the diagram shown in Fig.1(a) of the main text.
The corresponding coefficients are calculated as one-loop integral of four fermion Green's functions (Eq.~\eqref{eq:invG}),
\begin{equation}
  \begin{aligned}
   \beta_1&=\frac{T}{4\pi^2 v_F^2}\sum_n\int_{-E_F}^{E_F}\frac{dxdy}{(i\omega_n-x)^2(i\omega_n-y)^2},\\
    \beta_2&=\frac{T}{4\pi^2 v_F^2}\sum_n\int_{-E_F}^{E_F}\frac{-dxdy}{(i\omega_n-x)^2(\omega_n^2+y^2)},\\
     \beta&=\frac{T}{4\pi^2 v_F^2}\sum_n\int_{-E_F}^{E_F}\frac{dxdy}{(\omega_n^2+x^2)(\omega_n^2+y^2)}.
  \end{aligned}
\end{equation}
If we set $k_F=E_F/v_F=1$ and measure all the energies in units of $E_F$, then the above momentum integration and frequency summation can be done explicitly, leading to 
\begin{equation}
  \beta_1 \sim 1, ~~~\beta_2 \sim \ln\frac{1}{T}, ~~~\beta = \frac{1}{16 T}.\label{eq:betas}
\end{equation}

For the 6-leg and 8-leg diagrams, the leading contributions in terms of $E_F/T$ at each order are
\begin{widetext}
\begin{align}
&\gamma\left[\left(|\Psi_{\bQ}|^2+|\Psi_{-\bQ}|^2\right)|\Psi_{\bP}|^2|\Psi_{-\bP}|^2+ \left(|\Psi_{\bP}|^2+|\Psi_{-\bP}|^2\right)|\Psi_{\bQ}|^2|\Psi_{-\bQ}|^2+\left(\sum_{i=1}^4|\Psi_{\bQ_i}|^2\right)\left(\Psi_{\bP}^*\Psi_{-\bP}^*\Psi_{\bQ}\Psi_{-\bQ}+h.c.~\right)\right]\nonumber\\
    +&\frac{\zeta}{2}\left[\left(\Psi_{\bP}^*\Psi_{-\bP}^*\Psi_{\bQ}\Psi_{-\bQ}\right)^2+h.c.~\right]+4\zeta|\Psi_{\bQ}\Psi_{-\bQ}\Psi_{\bP}\Psi_{-\bP}|^2+\zeta\left(|\Psi_{\bP}|^2|\Psi_{-\bP}|^2+|\Psi_{\bQ}|^2|\Psi_{-\bQ}|^2\right)\left(\Psi_{\bP}^*\Psi_{-\bP}^*\Psi_{\bQ}\Psi_{-\bQ}+h.c.~\right)\nonumber\\
    +&\frac{\zeta}{4}\left(\sum_{i=1}^4|\Psi_{i}|^4\right)\left(\Psi_{\bP}^*\Psi_{-\bP}^*\Psi_{\bQ}\Psi_{-\bQ}+h.c.~\right)+\frac{\zeta}{2}\left(|\Psi_{\bQ}|^4+|\Psi_{-\bQ}|^4\right)|\Psi_{\bP}|^2|\Psi_{-\bP}|^2+\frac{\zeta}{2}\left(|\Psi_{\bP}|^4+|\Psi_{-\bP}|^4\right)|\Psi_{\bQ}|^2|\Psi_{-\bQ}|^2\nonumber\\
    +&\frac{\zeta}{2}\left(|\Psi_{\bP}|^2+|\Psi_{-\bP}|^2\right)\left(|\Psi_{\bQ}|^2+|\Psi_{-\bQ}|^2|\right)\left[|\Psi_{\bP}|^2|\Psi_{-\bP}|^2+|\Psi_{\bQ}|^2|\Psi_{-\bQ}|^2+2\left(\Psi_{\bP}^*\Psi_{-\bP}^*\Psi_{\bQ}\Psi_{-\bQ}+h.c.~\right)\right], \label{eq:8leg}
\end{align}
\end{widetext}
For compactness we have suppressed the momenta carried by each fields and the integration over conserved momenta.

Below we briefly discuss how to obtain \eqref{eq:8leg}.
For $n=3$, the internal Green's functions runs over the set $\{1,2,3,4,1,2\}$, where we denote $G_i$ by $i$ for brevity. Note that the set is equivalent to $\{1,2,3,4,2,3\}$, $\{1,2,3,4,3,4\}$ and $\{1,2,3,4,4,1\}$ by $D_4$ symmetry. Thus all of these sets should be considered, and they combined to give the $\gamma$ terms in Eq.~\eqref{eq:8leg}.

The results for $n=4$ contains more terms. First of all, all the fermion Green's functions can run over the set $\{1,2,3,4,1,2,3,4\}$. The set is invariant under $D_4$ group operation so this is a complete class. The momentum integral yields a factor of $\zeta$ for this class, and this is presented in the second line of Eq.~\eqref{eq:8leg}. 
The second class comes from the set $\{1,2,3,4,1,1,2,2\}$ and its symmetry related ones $\{1,2,3,4,2,2,3,3\}$, $\{1,2,3,4,3,3,4,4\}$ and $\{1,2,3,4,4,4,1,1\}$. The one-loop integral for these diagrams leads to a factor of $\zeta/4$, and this class corresponds to the third line of Eq.~\eqref{eq:8leg}. 
The last class comes from the set $\{1,2,3,4,1,1,2,4\}$, $\{1,2,3,4,2,2,1,3\}$, $\{1,2,3,4,3,3,2,4\}$ and $\{1,2,3,4,4,4,1,3\}$, for which the one-loop integral yields a factor of $\zeta/2$. This corresponds to the last line of Eq.~\eqref{eq:8leg}. It can be checked that all other sets doest not meet the momentum conservation, i.e. they will lead to some external legs for which the momentum is neither $\pm\bQ$ nor $\pm\bP$. Therefore, Eq.~\eqref{eq:8leg} is all the leading contributions at $\mathcal{O}(|\Psi_{\bQ_i}|^8)$.

Similarly, the coefficients of $\gamma$ and $\zeta$ are given by one-loop integration of six and eight Green's functions, namely
\begin{equation}
  \begin{aligned}
    \gamma&=T\sum_n\int \frac{dx dy}{4\pi^2}\left(\frac{1}{i\omega_n-x}\right)^2\left(\frac{1}{i\omega_n-y}\right)^2\frac{1}{i\omega_n+x}\frac{-1}{i\omega_n+y}\\
    &=\frac{1}{16T^3\pi^4}\sum_{n=-\infty}^{+\infty}\frac{1}{(2n+1)^4}\\
    &=\frac{1}{768T^3},
    \label{eq:gamma}
  \end{aligned}
\end{equation}
and 
\begin{equation}
  \begin{aligned}
    \zeta &= T\sum_n\frac{dx dy}{4\pi^2}\left(\frac{1}{i\omega_n-x}\right)^2\left(\frac{1}{i\omega_n-y}\right)^2\left(\frac{1}{i\omega_n+x}\right)^2\left(\frac{1}{i\omega_n+y}\right)^2\\
    &= \frac{1}{16 T^5\pi^6}\sum_{n=-\infty}^{+\infty}\frac{1}{(2n+1)^6}\\
    &=\frac{1}{7680T^5}.
    \label{eq:zeta}
  \end{aligned}
\end{equation}

\section{Details on decoupling the PDW fields} 
\label{sec:details_about_decoupling_the_pdw_fields}


The higher-order interactions presented in Eq.~\eqref{eq:8leg} can be greatly simplified by using composite fields including those corresponding to $4e$-SC
  \begin{align}
   \Psi_{\bP}^*\Psi_{-\bP}^*\Psi_{\bQ}\Psi_{-\bQ}+h.c. &= \frac{1}{2}|\Phi_s|^2-\frac{1}{2}|\Phi_d|^2, \nonumber\\
    |\Psi_{\bP}|^2|\Psi_{-\bP}|^2&=\frac{1}{4}|\Phi_s-\Phi_d|^2,\nonumber\\
    |\Psi_{\bQ}|^2|\Psi_{-\bQ}|^2&=\frac{1}{4}|\Phi_s+\Phi_d|^2,  \\
    \Psi_{\bQ}\Psi_{-\bQ}\Psi_{\bP}\Psi_{-\bP}&=\frac{1}{16}\(\Phi_s^2-\Phi_d^2\)\nonumber\\
   \! \! \! \!\left(|\Psi_{\bP}|^2+|\Psi_{-\bP}|^2\right)\left(|\Psi_{\bQ}|^2+|\Psi_{-\bQ}|^2|\right)&=  \frac{1}{{4}}  \(\phi^2 - \mathcal{N}^2\),\nonumber
     \end{align}
where in the last line $\phi\equiv |\Psi_{\bP}|^2+|\Psi_{-\bP}|^2+|\Psi_{\bQ}|^2+|\Psi_{-\bQ}|^2$ is the Gaussian fluctuation and $\mathcal{N}\equiv |\Psi_{\bP}|^2+|\Psi_{-\bP}|^2-|\Psi_{\bQ}|^2-|\Psi_{-\bQ}|^2$ is the nematic order. It is important to note that this re-expression of quartic combinations of $\Phi$ in terms of bilinear operators is not unique. For example, $|\Psi_{\bP}|^2|\Psi_{-\bP}|^2$ can also be rewritten as $\[(|\Psi_{\bP}|^2+|\Psi_{-\bP}|^2)^2-(|\Psi_{\bP}|^2-|\Psi_{-\bP}|^2)^2\]/4$, which does not explicitly depends on $4e$-SC flucuations.  However, the re-expression can be made unique after extending the $\Psi$ field to an $M$-component $\Psi^I$, e.g.,
\begin{align}
   |\Psi_{\bP}|^2|\Psi_{-\bP}|^2 &\to \frac{1}{M}\sum_{I,J =1}^{M} \Psi_{\bP}^I \Psi_{\bP}^{*J} \Psi_{-\bP}^I \Psi_{-\bP}^{*J}\equiv \frac{M}{4}|\Phi_s-\Phi_d|^2.\\
      \textrm{where } \Phi_{s,d} &=\frac{1}{M} \sum_I \(\Psi_{\bQ}^I \Psi_{-\bQ}^I\pm\Psi_{\bP}^I\Psi_{-\bP}^I\).
    \label{eq:14}
\end{align}
Such a large-$M$ extension justifies the mean-field theory for $4e$-SC, and thus different extension schemes correspond to different mean-field ansatze. 

We insert to the partition function the following $\delta$-function identities~\footnote{$\lambda_d$ and $\lambda_d^*$, as well as $\Delta_d$ and $\Delta_d^*$, are independent fields. The integration domains for $\lambda_d$, $\lambda_d^*$, and $\mu$ are along the imaginary axis, but their saddle points may not be.}
  \begin{align}
    1&\propto\int\mathcal{D}[\lambda_d,\Delta_d^*]\exp\left[-\frac{1}{T}\int_{\bq}\lambda_d(\bq)\({\Delta}_d^*(-\bq)-\Phi^*_d(\bq)\)\right]\nonumber\\
    1&\propto\int\mathcal{D}[\varphi,\mu]\exp\left[-\frac{1}{T}\int_{\bq} \mu(\bq)\(\varphi(-\bq)-\phi(-\bq)\)\right]
\label{eq:ID}
\end{align} 
and the saddle points for the $\lambda$'s and $\mu$ ensures that one can replace in Eq.~\eqref{eq:8leg} the bilinear fluctutations $\Phi_d$ and $\phi$ with local fields $\Delta_d$ and $\varphi$. The resulting theory is 
 \begin{align}
    &{F}[\Psi, \Delta_d, \varphi, \lambda_d, \mu] \nonumber\\
    &= \sum_{i} \int_{\bq} \[\alpha(T) + \kappa\bm q^2\] \left | \Psi_i(\bm q)\right |^2 -\frac{\beta}{2}\int_{\bq}|\Delta_d(\bq)|^2  -\frac{\gamma}{4}\int_{\bq,\bp}\Delta_d(\bq)\Delta_d^*(\bq+\bp)\phi(\bp)\nonumber\\
  &+\frac{\zeta}{16}\int_{\bq,\bk, \bp}\Delta_d(\bq)\Delta_d^*(\bq+\bp)\Delta_d(\bk)\Delta_d^*(\bk-\bp) -\frac{\zeta}{{16}}\int_{\bq,\bk, \bp}\Delta_d(\bq)\Delta_d^*(\bq+\bp)\varphi(\bk)\varphi(\bp-\bk)\nonumber\\
    &+\int_{\bq}[\lambda_d(\bq)\({\Delta}_d^*(\bq)-\Phi^*_d(\bq)\)+h.c.]+\int_{\bq}\mu(\bq)\(\varphi(-\bq)-\phi(-\bq)\)+ \mathcal{O}(\Psi^{10})
    \label{eq:15a}
  \end{align}
  As a mean-field ansatz, we consider saddle-point solutions of the composite fields  that are spatially uniform, with $\Delta_d(\bm q) = \Delta_{d}\delta (\bm q)$, and $\varphi(\bq) = \varphi \delta(\bq)$.
Using the regularization $\delta(\bq=0)=V$ where $V$ is the volume (area) of the system and defining the free energy density $\mathcal{F}\equiv F/V$, Eq.~\eqref{eq:15a} then becomes
\begin{align}
    \mathcal{F}[\Psi, \Delta_d, \varphi, \lambda_d, \mu )&= \sum_{i,\bq} \[\alpha(T) + \kappa\bm q^2 -\frac{\gamma}{4}|\Delta_d|^2-\mu\] \left | \Psi_i(\bm q)\right |^2-\frac{\beta}{2}|\Delta_d|^2 +\frac{\zeta}{16}|\Delta_d|^4
-\frac{\zeta}{{16}}|\Delta_d|^2\varphi^2\nonumber\\
&+\lambda_d\({\Delta}_d^*-\Phi^*_d\)+\lambda_d^*\({\Delta}_d-\Phi_d\)+\mu\varphi+ \mathcal{O}(\Psi^{10}),
    \label{eq:16}
  \end{align}
  where $\lambda_d\equiv \lambda_d(\bq =0)/V$, $\mu \equiv \mu(\bq =0)/V$, and we have used the fact that ${V}\int_{\bq}\equiv \sum_{\bq}$ in the continuum limit. 

\section{Details on mean-field theory for $4e$-SC}
\label{sec:mft_sm}
  To justify a mean-field theory, we formally extend Eq.~\eqref{eq:16} to a large-$M$ version using the scheme described in Eq.~\eqref{eq:14}. Together with rescaling $\lambda_d\to M\lambda_d$ and $\mu\to M\mu$, Eq.~\eqref{eq:16} becomes
\begin{align}
    \mathcal{F}[\Psi, \Delta_d, \varphi, \lambda_d, \mu )&= \sum_{J=1}^M\sum_{i}\int_{\bq} \[\alpha(T) + \kappa\bm q^2 -\frac{\gamma}{4}|\Delta_d|^2-\mu\] \left | \Psi_i^J(\bm q)\right |^2 -\frac{M\beta}{2}|\Delta_d|^2 +\frac{M\zeta}{16}|\Delta_d|^4
-\frac{M\zeta}{{16}}|\Delta_d|^2\varphi^2\nonumber\\
&+M\lambda_d\({\Delta}_d^*-\Phi^*_d\)+M\lambda_d^*\({\Delta}_d-\Phi_d\)+M\mu\varphi+ \mathcal{O}(\Psi^{10}).
\label{eq:s19SM} 
  \end{align}
Since $\mathcal{F}$ is quadratic in $\Psi$, we can integrate it out. We get, up to higher-order terms
\begin{align}
    \mathcal{F}( \Delta_d, \varphi, \lambda_d, \mu )/M =& {2}T\int_{\bq} \ln \[ (\alpha(T) + \kappa\bm q^2 -\frac{\gamma}{4}|\Delta_d|^2-\mu)^2 - |\lambda_d|^2\]\nonumber\\
  &-\frac{\beta}{2}|\Delta_d|^2 +\frac{\zeta}{16}|\Delta_d|^4
-\frac{\zeta}{{16}}|\Delta_d|^2\varphi^2+\lambda_d{\Delta}_d^*+\lambda_d^*{\Delta}_d+\mu\varphi ,\label{eq:23}
\end{align}
where the upper cutoff is a proper combination of high-energy scale $E_F$ and short distance scale $n^{-1/2}$, which in our units is 1. We have also used $\frac{1}{V}\sum{\bq}\equiv \int_{\bq}$. 

In the $M\to \infty$ limit, the partition function $Z=e^{-\mathcal{F}V/T}$ is completely determined by the saddle points of $\Delta_d, \varphi, \lambda_d$ and $\mu$.
One can take the saddle-point equations to eliminate the Lagrange multiplier fields $\lambda_d$ and $\mu$. We get from the requirements $\partial\mathcal{F}/\partial\varphi=0$, $\partial\mathcal{F}/\partial\lambda_d^*=0$ and  $\partial\mathcal{F}/\partial\mu=0$ that
\begin{align}
\mu=& \frac{\zeta}{{8}}|\Delta_d|^2\varphi \label{eq:22a}\\
|\lambda_d| =&r  \tanh \frac{{2}\pi \kappa|\Delta_d|}{T} \label{eq:22b} \\
\varphi = &\frac{T}{\pi \kappa} \ln\(\frac{{\kappa}}{r} \cosh \frac{ {2}\pi \kappa |\Delta_d|}{T} \),  \label{eq:22}\\
\textrm{where } r\equiv & \alpha(T)-\frac{\gamma}{{4}}|\Delta_d|^2-\frac{\zeta}{{8}}|\Delta_d|^2\varphi, \label{eq:22c}
\end{align}
{where to obtain Eq.~\eqref{eq:22} we have used Eq.~\eqref{eq:22b}.} Note that in order to legitimize the procedure of integrating out PDW orders, we need to keep $r$ as positive definite. This will be justified in our following calculation for $\Delta_d$.  
Eqs.~(\ref{eq:22a}, \ref{eq:22b}) can be used to eliminate the Lagrange multiplier fields $\lambda$ and $\mu$ in Eq.~\eqref{eq:23}, which yields 
\begin{align}
    &\frac{\mathcal{F}( \Delta_d )}{M}=-\frac{\beta}{2}|\Delta_d|^2 +\frac{\zeta}{16}|\Delta_d|^4 +\frac{\zeta}{16}|\Delta_d|^2\varphi^2+r\varphi+\frac{rT}{\pi\kappa} \nonumber\\   
     &=\alpha(T)\varphi-\(\frac{\beta}{2}+\frac{T\gamma}{4\pi \kappa}\)|\Delta_d|^2 -\(\frac{\gamma}{4}+\frac{T\zeta}{8\pi \kappa}\)|\Delta_d|^2\varphi -\frac{\zeta}{16}|\Delta_d|^2\varphi^2+\frac{\zeta}{16}|\Delta_d|^4 + \rm{const},
     \label{eq:28c}
\end{align}
where at the second equality we have used Eq.~\eqref{eq:22c} to eliminate $r$.

The $\varphi$ dependence in Eq.~\eqref{eq:28c} can be further eliminated using Eq.~\eqref{eq:22}, which does not admit a closed-form solution but can be expressed in Taylor expansion
\begin{equation}
  \varphi=\frac{T}{\pi\kappa}\ln\frac{\kappa}{\alpha}+c_2|\Delta_d|^2+c_4|\Delta_d|^4+c_6|\Delta_d|^6+...\label{eq:phiexpand}
\end{equation}
where the first term is the familiar logarithmic divergence of Gaussian fluctuations in 2d, and the rest is the back action from $4e$-SC fluctuations. 
Substituting this expansion into Eq.~\eqref{eq:22} and matching the coefficients on both sides, we find, omitting high-order terms in $c_6$ due to space, 
\begin{equation}
  \begin{aligned}
    c_2&=\frac{2 \pi  \kappa }{T}+\frac{\gamma  T}{4 \pi  \alpha  \kappa }+ \frac{\zeta  T^2 }{8 \pi ^2 \alpha  \kappa ^2}\ln\frac{\kappa }{\alpha },\\
    c_4&=-\frac{4 \pi ^3 \kappa ^3}{3 T^3}+\frac{\zeta }{4 \alpha }+\frac{\gamma ^2 T}{32 \pi  \alpha ^2 \kappa }+\frac{\gamma  \zeta  T^2}{32 \pi ^2 \alpha ^2 \kappa ^2} \ln \frac{\kappa }{\alpha }+\frac{\zeta ^2 T^3 }{64 \pi ^3 \alpha ^2 \kappa ^3}\ln \frac{\kappa }{\alpha }+\frac{\zeta ^2 T^3 }{128 \pi ^3 \alpha ^2 \kappa ^3}\ln ^2\frac{\kappa }{\alpha }\\
        c_6&=\frac{64 \pi ^5 \kappa ^5}{45 T^5} -\frac{\pi ^2 \zeta  \kappa ^2}{6 \alpha  T^2}+ \cdots
  \end{aligned}
\end{equation}

  Next we integrate out the PDW fields, and replace $\varphi$, $\lambda_d$ and $\mu $ with their saddle point values. This leads to the following GL free energy for $\Delta_d$ only,
  \begin{equation}
  {\mathcal{F}( \Delta_d)}/{M} = A(T) |\Delta_d|^2+B(T) |\Delta_d|^4  + C(T)|\Delta_d|^6 + \cdots,
\end{equation}
where the GL coefficients are
  \begin{align}
    A(T)&=\frac{2 \pi  \alpha  \kappa }{T}-\frac{\beta }{2}-\frac{\gamma  T }{4 \pi  \kappa }\ln\frac{\kappa }{\alpha }-\frac{\zeta  T^2 }{16 \pi ^2 \kappa ^2}\ln ^2\frac{\kappa }{\alpha },\nonumber\\
    B(T)&=-\frac{4 \pi ^3 \alpha  \kappa ^3}{3 T^3}-\frac{\pi  \gamma  \kappa }{2 T}+\frac{\zeta }{16}-\frac{\gamma ^2 T}{32 \pi  \alpha  \kappa }-\frac{\zeta}{4}\left(\frac{\gamma  T^2 }{8 \pi ^2 \alpha  \kappa ^2}+1\right)  \ln \frac{\kappa }{\alpha }-\frac{\zeta ^2 T^3 }{128 \pi ^3 \alpha  \kappa ^3}\ln ^2\frac{\kappa }{\alpha },\nonumber\\
    C(T)&=\frac{64 \pi ^5 \alpha  \kappa ^5}{45 T^5}+\frac{\pi ^3 \gamma  \kappa ^3}{3 T^3}+\frac{\pi^2\zeta\kappa^2}{6T^2}\left(\ln\frac{\kappa}{\alpha}-\frac{3}{2}\right)+\cdots \label{eq:abc}
  \end{align}
 Using Eqs.~(\ref{eq:betas}, \ref{eq:gamma}, and \ref{eq:zeta}), we see that  all terms in Eq.~\eqref{eq:abc} are actually organized in increasing powers of $1/\kappa T\ll 1$ or $1/\kappa\alpha$. Assuming $\kappa\alpha\gtrsim 1$ as we will justify below, the coefficients can be greatly simplified. Indeed, we have already omitted many terms in higher powers of $1/\kappa T$ in $C(T)$. 
More importantly, higher-order terms in the PDW field $\Psi$  in Eq.~\eqref{eq:16} in fact enter all coefficients $A$, $B$, and $C$, but it is straightforward to see from the pattern that they are further suppressed by $1/\kappa T$, which thereby serves as a control parameter for the perturbative expansion of the free energy in terms of the PDW fields $\Psi$. In the opposite limit $\kappa T\ll 1$, the effective theory for PDW bosons becomes strongly coupled, and a perturbative expansion in $\Psi$ becomes inadequate.
  Keeping only the leading order contributions in the limit of $1/\kappa T, 1/\kappa\alpha \ll 1$, we arrive at Eq.(12) in the main text. In this limit, the leading contribution to $\mathcal{F}(\Delta_d)$ comes from quadratic and quartic terms of the PDW free energy.

\section{Anisotropy of PDW stiffness and suppression of vestigial CDW order} 
\label{sec:calculation_of_}

\begin{figure}
  \includegraphics[width=8cm]{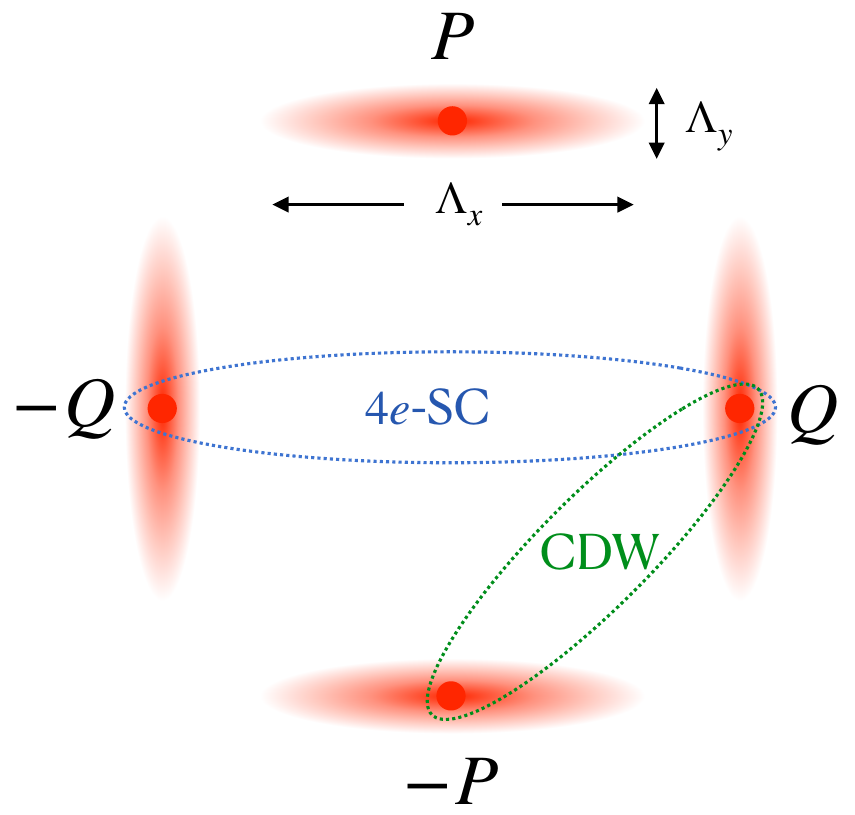}
  \caption{CDW order is suppressed by non-perfect nesting. In practice, the momentum integration for CDW is restricted to only a tiny area to retain the CDW momentum, which shrinks the available phase space and hence reduces the outcomes after integration.}\label{fig:nesting}
\end{figure}

From Eq.~\eqref{eq:betas} we see that at low temperatures, we only need to keep the $\beta$ term, which is much larger than $\beta_1$ and $\beta_2$. But even with only this term, there are different ways of decomposing it which result in different vestigial orders. As we have discussed in the main text that the charge-$4e$ channel is the leading channel, while the CDW channel can be neglected. This is true when the primary PDW susceptibility has some anisotropy as shown in \ref{fig:nesting}, which is in general the case with $C_4$ symmetry (See also Fig.1 in our main text). {For systems with continuous rotation symmetry, the anisotropy at any given PDW ordering momentum is even stronger -- in one direction the inverse susceptibility is quadratic in momentum, while it is quartic in the perpendicular direction.}

{As can be seen in \ref{fig:nesting}, the red regions indicates the momentum range of soft (low-energy) PDW fluctuations, which are anisotropic around each PDW order momenta.  The $4e$-SC involves low-energy PDW bosons with opposite momenta, such that as long as one of the boson is low-energy, so does the other. However, the CDW order involves bosons that differ in momentum by, say $\bm P+\bm Q$. In order for both bosons to be low-energy, the momentum of one PDW boson should be near $\pm \bm Q$ and the other near $\mp \bm P$. Compared with that for $4e$-SC, the effective momentum range for bosons contributing to the CDW instability is much smaller, thus leading to a much weaker tendency toward CDW. As we mentioned in the main text, this is similar to the argument in fermionic systems, in which CDW instabilities are weaker due to the general lack of nesting of the fermionic dispersion.~\cite{PhysRevB.90.035149}. }

{By  analyzing their free energy separetly, we can show quantitatively that the $4e$-SC order always has a higher transition temperature than CDW. We note that the same four-boson interaction can be decomposed in either the particle-particle channel or the particle-hole channel, i,e,}
\begin{align}
\textrm{$4e$-SC channel:~~}  \beta&\int _{\bm{q}_1,\bm{q}_2,\bm{q}_3}\left[\Psi_{\bm{P}}(\bm{q}_1)\Psi_{-\bm{P}}(-\bm{q}_1+\bm{q}_3)\right]\left[\Psi^*_{\bm{Q}}(\bm{q}_2)\Psi^*_{-\bm{Q}}(-\bm{q}_2+\bm{q}_3)\right]+h.c..\label{eq:4e} \\
 \textrm{CDW channel:~~}      \beta&\int_{\bm{q}_1,\bm{q}_2,\bm{q}_3}\left[\Psi_{\bm{P}}(\bm{q}_1)\Psi^*_{\bm{Q}}(\bm{q}_1+\bm{q}_3)\right]\left[\Psi_{-\bm{P}}(\bm{q}_2)\Psi^*_{-\bm{Q}}(\bm{q}_2-\bm{q}_3)\right]+h.c..\label{eq:cdw}
\end{align}
{Moreover, the interaction can be made equally attractive in $4e$-SC and CDW channels by choosing the proper sign-changing patterns in momentum space.}

{Due to the attractive interaction, one can separately introduce CDW and $4e$-SC order parameters that couple to PDW bosons via the bilinear term. Separately introducing the corresponding auxiliary fields to obtain the free energies for CDW and $4e$-SC, extending them to large-$M$, and integrating out the PDW field while keeping the anisotropy of stiffness around each PDW wave vector, we have for $4e$-SC order $\Delta_d$,
\begin{align}
    &\mathcal{F}_{\rm 4e}( \Delta_d, \varphi, \lambda_d, \mu )/M \nonumber\\
    =& {2}T\int_{\bq} \ln  \det\begin{pmatrix}
    \kappa_1 q_x^2 + \kappa_2 q_y^2+r(\Delta_d,\mu) & 0 & -\lambda_d & 0\\
    0 &  \kappa_1 q_y^2 + \kappa_2 q_x^2 +r(\Delta_d,\mu) & 0 & \lambda_d\\
    -\bar\lambda_d & 0 &  \kappa_1 q_x^2 + \kappa_2 q_y^2+r(\Delta_d,\mu) & 0\\
    0 & \bar\lambda_d & 0 & \kappa_1 q_y^2 + \kappa_2 q_x^2+r(\Delta_d,\mu)
  \end{pmatrix}
    \nonumber\\
  &-\frac{\beta}{2}|\Delta_d|^2 +\frac{\zeta}{16}|\Delta_d|^4
-\frac{\zeta}{{16}}|\Delta_d|^2\varphi^2+\lambda_d{\Delta}_d^*+\lambda_d^*{\Delta}_d+\mu\varphi +\cdots ,\label{eq:23a}
\end{align}
while for CDW order $\Upsilon_d$
\begin{align}
    &\mathcal{F}_{\rm CDW}( \Upsilon_d, \varphi, \lambda_d, \mu )/M\nonumber\\
     =& {2}T\int_{\bq} \ln  \det\begin{pmatrix}
    \kappa_1 q_x^2 + \kappa_2 q_y^2+r(\Upsilon_d,\mu) & \lambda_d & 0 & 0\\
    \bar\lambda_d &  \kappa_1 q_y^2 + \kappa_2 q_x^2 ++r(\Upsilon_d,\mu) & 0 & 0\\
    0 & 0 &  \kappa_1 q_x^2 + \kappa_2 q_y^2+r(\Upsilon_d,\mu) & -\lambda_d\\
    0 & 0 & -\bar\lambda_d & \kappa_1 q_y^2 + \kappa_2 q_x^2+r(\Upsilon_d,\mu)
  \end{pmatrix}
    \nonumber\\
  &-\frac{\beta}{2}|\Upsilon_d|^2 +\frac{\zeta}{16}|\Upsilon_d|^4
-\frac{\zeta}{{16}}|\Upsilon_d|^2\varphi^2+\lambda_d{\Upsilon}_d^*+\lambda_d^*{\Upsilon}_d+\mu\varphi + \cdots. \label{eq:23b}
\end{align} 
Here in general $\kappa_1\neq \kappa_2$, as can be seen from \ref{fig:nesting}, and $r(\Delta, \mu) = \alpha(T) - \frac{\gamma}{4}|\Delta|^2 - \mu$. Note that only difference between $\mathcal{F}_{\rm CDW}$ and $\mathcal{F}_{\rm 4e}$ is in the off-diagonal matrix elements. }

{By evaluating the determinants of the $4\times 4$ matrices,  it is straightforward to see that for $\mathcal{F}_{\rm 4e}$, the anisotropy in $\kappa_{1,2}$ can be absorbed into a redefinition of $q_x$ and $q_y$, leading to an effective $\kappa_{\rm eff}= \sqrt{\kappa_1\kappa_2}$ in place of the $\kappa$ used in the main text and in the previous sections. }

{However, the same cannot be done for $\mathcal{F}_{\rm CDW}$, due to the mismatch of the diagonal terms in the $2\times 2$ blocks of the matrix. Due to this mismatch, evaluating the two determinants and using simple algebraic relations, we can prove that for the same arguments with $\lambda_d\neq 0$
\be
\mathcal{F}_{\rm CDW}( \Delta, \varphi, \lambda_d, \mu )>\mathcal{F}_{\rm 4e}( \Delta, \varphi, \lambda_d, \mu ).
\label{eq:comp}
\ee}

{From Eq.~(\ref{eq:22a}-\ref{eq:22}), both free energy functions admit a trivial solution $\Delta_d=\Upsilon_d=0$, $\lambda_d=0$, $\mu=0$, $\varphi = ({T}/{\pi \kappa_{\rm eff}})\ln {\kappa}/{\alpha}$. At these values $\mathcal{F}_{\rm 4e}=\mathcal{F}_{\rm CDW}$, which we define as $\mathcal{F}_0$. The first order phase transition occurs for $4e$-SC when at a nonzero saddle-point value of $\Delta_{\rm 4e}$, $ \mathcal{F}_{\rm 4e}( \Delta_d, \varphi, \lambda_d, \mu ) = \mathcal{F}_0$, such that the trivial solution is no longer the global minimum. From Eq.~\eqref{eq:comp}, the saddle point value of $\mathcal{F}_{\rm CDW}( \Upsilon_d, \varphi, \lambda_d, \mu )$ should be larger than $\mathcal{F}_0$, thus indicating the CDW transition has not yet occured. This completes our proof that $4e$-SC has a higher transition temperature than the vestigial CDW order.}

\section{Equivalence between HS transformation and the method of Lagrangian multipliers} 
\label{sec:equi}

In a special case when $\gamma$ and $\zeta$ terms are absent from the action, one can alternatively perform HS (HS) transformation to analyze the charge-$4e$ order. Below we show that the HS transformation yields exactly the same results as we obtain by using Lagrangian multiplier.  The action we consider is 
\begin{equation}
  \int \frac{d^2q}{4\pi^2}\left[ (\kappa q^2+\alpha)\sum_{i=1}^4|\Psi_{\bQ_i}|^2\right]-\frac{\beta}{2}|\Phi_d|^2
\end{equation}
with $\beta>0$. After HS transformation, the action becomes 
\begin{equation}
  \frac{\beta}{2}\left(|\Delta_d|^2-\Delta_d{\Phi_d}^*-\bar{\Delta}_d\Phi_d\right)+ \int \frac{d^2q}{4\pi^2}\left[ (\kappa q^2+\alpha)\sum_{i=1}^4|\Psi_{\bQ_i}|^2\right]
\end{equation}
Again one can integrate out the PDW order parameters using the basis $\Psi=(\Psi_{\bQ},\Psi_{\bP},\Psi^\dagger_{-\bQ},\Psi^\dagger_{-\bP})^T$. This leads to the following effective action for the charge-$4e$ order
\begin{equation}
  \frac{\beta}{2}\Delta_d^2+ T\int \frac{d^2q}{4\pi^2}\text{tr}\ln A(q)\label{eq:act2}
\end{equation}
where 
\begin{equation}
  A(q)=
  \begin{pmatrix}
    \kappa q^2+\alpha & 0 & -\frac{\beta}{2}\Delta_d & 0\\
    0 &  \kappa q^2+\alpha & 0 & \frac{\beta}{2}\Delta_d\\
    -\frac{\beta}{2}\bar{\Delta}_d & 0 &  \kappa q^2+\alpha & 0\\
    0 & \frac{\beta}{2}\bar{\Delta}_d & 0 &  \kappa q^2+\alpha
  \end{pmatrix}.
\end{equation}
Performing the tr~$\ln$ operation, Eq.~\eqref{eq:act1} becomes
\begin{equation}
  \frac{\beta}{2}\Delta_d^2+ T\int \frac{d^2q}{2\pi^2}\ln\left[( \kappa q^2+\alpha)^2-\frac{\beta^2}{4}\Delta_d^2\right]
\end{equation}
The saddle point of this action reads
\begin{equation}
  \begin{aligned}
    1&={\beta}T\int\frac{d^2q}{4\pi^2}\frac{1}{( \kappa q^2+\alpha)^2-\frac{\beta^2}{4}\Delta_d^2}\\
    &=\frac{\beta T}{4\pi\kappa}\frac{1}{\frac{\beta}{2}\Delta_d}\text{arccoth}\frac{\alpha}{\frac{\beta}{2}\Delta_d},
  \end{aligned}
\end{equation}
or equivalently, 
\begin{equation}
  \Delta_d=\frac{T}{2\pi\kappa}\text{arccoth}\frac{2\alpha}{\beta\Delta_d}~~\text{or}~~\tanh(2\pi\kappa \Delta_d/T)=\frac{\beta\Delta_d}{2\alpha}. \label{eq:HSresult}
\end{equation}
In obtaining the above saddle point equations, we have used the condition that $\alpha>\frac{\beta}{2}\Delta_d$, as is required for the convergence of the Gaussian integral in order to integrate out PDW fields. Eq.~\eqref{eq:HSresult} is the results obtained via HS transformation at quartic level. 

To compare Eq.~\eqref{eq:HSresult} with the one obtained using the method of Lagrangian multiplier, we note  that in the absence of $\gamma$ and $\zeta$, we have [see Eq.~\eqref{eq:22}]
\begin{equation}
  \varphi = \frac{T}{\pi \kappa} \ln\(\frac{{\kappa}}{\alpha} \cosh \frac{ {2}\pi \kappa |\Delta_d|}{T} \).
\end{equation}
The free energy for $\Delta_d$ then becomes 
\begin{align}
 \!\!   \!\!   \!\!  \!\!  \! \! \!  \!  \mathcal{F}[\Delta_d]=-\frac{\beta}{2}|\Delta_d|^2+\frac{\alpha T}{\pi\kappa}\left[1+\ln\(\frac{{\kappa}}{\alpha} \cosh \frac{ {2}\pi \kappa |\Delta_d|}{T} \)\right].
\end{align}

Variation of this free energy with respect to $|\Delta_d|$ and set it to zero, we immediately obtain the gap equation
\begin{equation}
  -\frac{\beta}{2}\Delta_d+\alpha\tanh(2\pi\kappa \Delta_d/T)=0
\end{equation}
which is exactly the same as Eq.~\eqref{eq:HSresult}. Therefore, we have proved that at the level of $\mathcal{O}(\Psi^4)$, both the HS transformation and the method of Lagrangian multiplier yield the same result. However, the HS transformation, which is built on Gaussian integral, fails to work once higher order terms of the primary order parameters are included, but the Lagrangian multiplier method is still valid.

\bibliography{charge4e}